\documentclass[11pt, a4paper]{article}
\pdfoutput=1

\usepackage{jheppub}    

\usepackage{slashed}    
\usepackage{tikz}
\usepackage{tabularx}
\usepackage{booktabs}

\usetikzlibrary{patterns}
\usetikzlibrary{decorations.pathreplacing}

\newcommand{\pdiff}[2]{\frac{\partial #1}{\partial #2}}

\title{
Identifying the production process of new physics at colliders; symmetric or asymmetric?
}

\author[a,b]{Sung Hak Lim}

\affiliation[a]{Department of Physics, KAIST, 291, Daehak-ro, Yuseong-gu, Daejeon, 34141, Korea}
\affiliation[b]{Center for Theoretical Physics of the Universe, Institute for Basic Science (IBS), 34051, KAIST Munji Campus Faculty Wing 3rd Floor, 193, Munji-ro, Yuseong-gu, Daejeon, Korea}

\emailAdd{sunghak.lim@kaist.ac.kr}

\preprint{CTPU-16-06}

\abstract{
We propose a class of kinematic variables, which is a smooth generalization of min-max type mass variables such as the Cambridge-$M_{T2}$ and $M_2$, for measuring a mass spectrum of intermediate resonances in a semi-invisibly decaying pair production. 
While kinematic endpoints of min-max type mass variables are only sensitive to a heavier resonance mass, kinematic endpoints of new variables are sensitive to all masses.
These new mass variables can be used to resolve a mass spectrum, so that if the true mass spectrum is asymmetric, then the kinematic endpoints are separate while the endpoints are the same for the symmetric true mass spectrum. 
We demonstrate the behavior of kinematic endpoint of these new variables in pair production of two-body and three-body decays with one invisible particle.
}

\keywords{Hadron-Hadron scattering, spectroscopy, particle and resonance production, beyond Standard Model, supersymmetry}

\arxivnumber{1603.01981}

\begin{document}
\maketitle
\flushbottom

%
\section{Introduction}
%
Uncovering resonances hiding behind events with multiple invisible particles at a collider is essential for the discovery of new physics. 
Events with missing energy are typically favored in popular new physics scenarios with dark matter candidates \cite{Feng:2010gw}, such as the weak scale supersymmetry \cite{Nilles:1983ge, Haber:1984rc} and the extra-dimensional model \cite{Appelquist:2000nn}. 
These new physics scenarios have conserved discrete symmetries, like $R$-parity and $KK$-parity, stabilizing the dark matter and leaving, at least, a pair of invisible particles in the new physics event. 
While the most straightforward approach to discovering intermediate resonances is finding out Breit-Wigner resonance peaks in a reconstructed invariant mass distribution, however, the resonance peak cannot be reconstructed in such new physics events with invisible particles.

Instead of reconstructing resonance peaks, kinematic singularities \cite{Kim:2009si}, such as endpoints of kinematic variables \cite{Hinchliffe:1996iu, Hinchliffe:1999zc, Tovey:2008ui, Han:2009ss, Barr:2010zj, Cho:2012er, Kim:2015bnd}, are often used to find out the new resonances.
For semi-invisibly decaying pair productions, kinematic endpoints of \emph{min-max type mass variables}, such as the Cambridge-$M_{T2}$ variable \cite{Lester:1999tx, Barr:2003rg, Cho:2007qv, Cheng:2008hk}, $M_2$ variable \cite{Barr:2011xt, Mahbubani:2012kx, Cho:2014naa, Cho:2015laa} and their variants \cite{Konar:2009wn, Cho:2009ve, Konar:2015hea}, are well-known to provide relevant information for measuring intermediate resonant particle masses as well as invisible particle masses \cite{Cho:2007qv, Gripaios:2007is, Barr:2007hy, Cho:2007dh, Konar:2009qr}.
The min-max type mass variables infer the momentum of invisible particles from a minimization of the \emph{maximum} of two interested resonance masses over physical momenta configurations consistent with transverse momentum conservation. 
By construction, these mass variables are bounded above by maximum of the true resonance masses \cite{Nojiri:2008hy, Barr:2009wu}.
The min-max type mass variables are useful especially when the masses of the two resonant mother particles are identical, i.e. symmetric, because kinematic endpoints of the mass variables saturate this upper bound. 
Hence, we can adapt the kinematic endpoints to reveal the true resonance mass \cite{Cho:2008cu, Aaltonen:2009rm, Chatrchyan:2013boa}.

However, kinematic endpoints of these min-max type mass variables have foibles when the true mass spectrum is asymmetric, i.e. non-identical masses. 
First, the upper bound only provide information on the heavier resonance mass \cite{Nojiri:2008hy, Barr:2009wu} even though two resonances construct the mass variables.
Trying to be assumption independent; indeed, we should identify the lighter resonance mass by kinematics.
Second, the kinematic endpoint often saturates the upper bound only when a considerable upstream momentum and a large total center-of-mass energy $\sqrt{\hat{s}}$ of colliding partons are supplied \cite{Mahbubani:2012kx, Barr:2009jv}.
Without such extreme kinematic conditions, the endpoint should be interpreted cautiously.
Therefore, we need special treatment for studying asymmetric mass spectrum within a framework of min-max type mass variables. 
Studies of generalized $M_{T2}$ with a hypothetical ratio of parents masses \cite{Barr:2009jv} and subsystem $M_{T2}$ for gluino-squark co-production \cite{Nojiri:2008vq} are examples of the non-trivial treatment.

In this regard, we introduce a broader class of mass variables as a generalization of the min-max type mass variables. 
Kinematic endpoints of these new variables depend on both masses of intermediate resonances and hence 
it can constrain two masses much more efficiently. 
While the mass measurement by the standard $M_{T2}$ endpoint in an asymmetric mass spectrum requires the extreme kinematic conditions that we have mentioned above, our newly introduced method can enhance the sensitivity by reducing the dependency of the extremal kinematic configurations.
We demonstrate kinematic endpoints behavior of these new variables in the case of a general asymmetric mass spectrum. 
After then,  We will also compare the key features of our new variables and the other extenstions of $M_{T2}$ briefly.\\

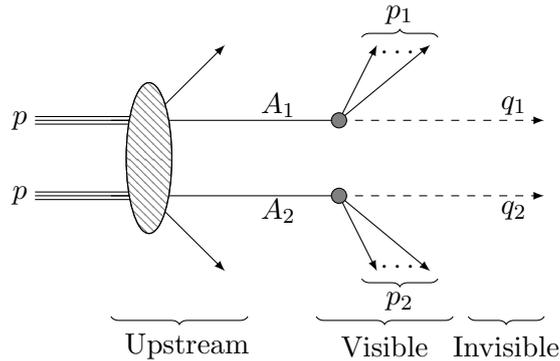
\begin{figure}
\begin{center}
\begin{tikzpicture}
\draw (-0.5,0.5) -- (-2,0.5);
\draw (-0.5,0.55) -- (-2,0.55);
\draw (-0.5,0.45) -- (-2,0.45);
\node[draw=none] at (-2.2,0.5) {$p$};
\draw (-0.5,-0.5) -- (-2,-0.5);
\draw (-0.5,-0.55) -- (-2,-0.55);
\draw (-0.5,-0.45) -- (-2,-0.45);
\node[draw=none] at (-2.2,-0.5) {$p$};

\draw[->,>=latex] (-0.5,0.5) -- (0.5,1.5);
\draw[->,>=latex] (-0.5,-0.5) -- (0.5,-1.5);

\draw (-1,0.5) -- (2,0.5);
\draw (-1,-0.5) -- (2,-0.5);
\node[draw=none] at (1.2,0.7) {$A_1$};
\node[draw=none] at (1.2,-0.7) {$A_2$};

\draw[->,>=latex,dashed] (2,0.5) -- (4.7,0.5);
\draw[->,>=latex,dashed] (2,-0.5) -- (4.7,-0.5);
\node[draw=none] at (4.3,0.7) {$q_1$};
\node[draw=none] at (4.3,-0.7) {$q_2$};

\draw[->,>=latex] (2,0.5) -- (2.5,1.5);
\draw[->,>=latex] (2,-0.5) -- (2.5,-1.5);
\draw[->,>=latex] (2,0.5) -- (3.2,1.5);
\draw[->,>=latex] (2,-0.5) -- (3.2,-1.5);

\draw[decorate,decoration={brace,amplitude=3pt}] 
    (2.3, 1.6) -- 
    (3.3, 1.6) ; 
\draw[decorate,decoration={brace,amplitude=3pt,mirror}] 
    (2.3, -1.6) -- 
    (3.3, -1.6) ; 
\node[draw=none] at (2.8,1.9) {$p_1$};
\node[draw=none] at (2.8,-1.9) {$p_2$};
\node[draw=none] at (2.8,1.4) {$\cdots$};
\node[draw=none] at (2.8,-1.45) {$\cdots$};

\filldraw [color=black, fill=gray] (2,0.5) circle (0.1);
\filldraw [color=black, fill=gray] (2,-0.5) circle (0.1);

\node[draw=none] at (0.0,-2.5) {Upstream};
\draw[decorate,decoration={brace,amplitude=3pt,mirror}] 
    (-1, -2.1) -- 
    (0.8,-2.1) ; 

\filldraw [color=black, fill=white] (-0.5,0) ellipse (0.3 and 1);
\draw[pattern=north west lines, pattern color=gray] (-0.5,0) ellipse (0.3 and 1);

\node[draw=none] at (2.6,-2.5) {Visible};
\draw[decorate,decoration={brace,amplitude=3pt,mirror}] 
    (1.7, -2.1) -- 
    (3.5,-2.1) ; 
    
\node[draw=none] at (4.2,-2.5) {Invisible};
\draw[decorate,decoration={brace,amplitude=3pt,mirror}] 
    (3.7, -2.1) -- 
    (4.7,-2.1) ;  
\end{tikzpicture}
\end{center}
\caption{\label{fig:process_topology} A typical pair production of two resonances $A_1$ and $A_2$ having masses $m_{A_1}$ and $m_{A_2}$ decaying semi-invisibly in a collider. 
$p_i$ is a sum of momenta of the visible particles from the resonance $A_{i}$. $q_i$ is a momentum of the invisible particle from the resonance $A_i$. A system having semi-invisibly decaying $N$ resonances can be considered in a similar fashion. }
\end{figure}

%
\section{Minimized Power Means}
%
Consider $N$ resonances with masses $\{m_{A_i}\}_{i = 1 \cdots N}$ described in figure \ref{fig:process_topology} for $N=2$. 
Since momenta of invisible particles $q_i$ are unknown, we need a guess to construct resonance masses $\{M_{A_i}\}_{i = 1 \cdots N}$. 
Trial resonance masses are defined by
\begin{equation}
M_{A_i}^2 (p_i^\mu, q_i^\mu) = \left( p_i + q_i \right)^2 \;.
\end{equation}
where $p_i$ is a sum of momenta of the visible particles from the resonance $A_{i}$. 
We should keep in mind that, this trial resonance mass is defined ambiguously when some visible particles from different resonances are reconstructed into the same type of objects. 
For example, in gluino pair production, all visible particles are reconstructed into jets, and correctly assigning each jet to their mother resonance is not an easy problem.  
In $M_{T2}$ analysis, to reduce a side effect of this combinatorial ambiguity, several methods are proposed \cite{Nojiri:2008hy,Lester:2007fq,Alwall:2009zu,Baringer:2011nh,Choi:2011ys,Curtin:2011ng}. 
Solving the combinatorial ambiguity is not the main focus of this paper, so we will assume that all the visible particles are assigned correctly to their mother resonance.

Then, we need an objective function of the trial resonance masses, to be minimized over the invisible degree of freedoms in the event.
As a smooth generalization of the maximum function, which is the objective function to be minimized in the min-max type mass variables, we consider \emph{power means of two resonance masses}. A \emph{power mean} with an exponent $p$ of $N$ arguments $\{x_i\}_{i = 1 \cdots N}$ is defined by
\begin{equation}
\label{eqn:definition_power_mean}
\hat{\mu}_p ( x_1, \cdots, x_N ) 
=
\left( \frac{1}{N} \sum_{i=1}^N (x_i)^p  \right)^{\frac{1}{p}} \;.
\end{equation}
This power mean interpolates the relevant scales monotonically from the maximum of the arguments to the minimum via the Pythagorean means, which are listed in Table \ref{tab:power_mean}.

\begin{table}[h]
\newcolumntype{L}{>{\raggedright\arraybackslash}X}
\begin{center}
\begin{tabularx}{0.8\textwidth}{cLcL}
\toprule
$p$ & \multicolumn{1}{c}{$\hat{\mu}_p$} & $p$ & \multicolumn{1}{c}{$\hat{\mu}_p$} \\
\cmidrule(r){1-2}\cmidrule(r){3-4}
$\infty$ & Maximum & 0  & Geometric mean \\
2        & Root mean square     & -1 & Harmonic mean \\
1        & Mean    & -$\infty$ & Minimum \\
\bottomrule
\end{tabularx}
\end{center}
\caption{\label{tab:power_mean} The power mean interpolates functions between a maximum and a minimum function including Pythagorean means as some special cases.}
\end{table}

To construct a mass-bounding variable using the power mean, we define a \textit{minimized power mean} by
\begin{equation}
\label{eqn:definition_minimized_power_mean}
\mu_{p,N} (\tilde{m}) = 
 \min_{\substack{
	\{ \mathbf{q}_i | i = 1 ,\cdots, N \}  \\
    { \sum_i q_i^a = \slashed{p} {}_T^a  }
}}   \hat{\mu}_p (M_{A_1} \left( p_1^\mu, q_1^\mu(\tilde{m}) \right),\cdots ,
M_{A_N}(p_N^\mu, q_N^\mu(\tilde{m}) ) ) \;,
\end{equation}
where $a = 1, 2$ denotes an index of transverse component, $q_i^\mu(\tilde{m})$ is a trial momentum of the invisible particle with a spatial momentum $\mathbf{q}_i$ and a trial mass $\tilde{m}$ from the resonance $A_i$, and $\slashed{p} {}_T^a$ is a missing transverse momentum. We note that $\mu_{p,N}$ is a function of trial masses $\tilde{m}$ of invisible particles. Trying to be more general, we can set the trial masses of each invisible particle differently like $M_{T2}$'s case \cite{Konar:2009qr,Barr:2009jv, Bai:2012gs}. 
Nevertheless, we will focus on invisible particles having identical masses, because the typical dark matter motivated new physics scenarios prefer various new particles decaying into the same type of particle, which is the dark matter.

The minimized power mean $\mu_{p,N}$ is an increasing function of $p$ like $\hat{\mu}_p$ in eq. (\ref{eqn:definition_power_mean}). For a given event, if $p < q$, then the $\mu_{p,N}$ satisfies the following inequality, 
\begin{equation}
\label{eqn:minimized_power_mean_inequality}
\mu_{p,N} \leq \mu_{q,N} \; .
\end{equation}
At the invisible momentum solution for $\mu_{q,N}$, $\hat{\mu}_p \leq \hat{\mu}_q = \mu_{q,N}$ because $\hat{\mu}_p$ is an increasing function of $p$. Since $\mu_{p,N}$ is a minimum of $\hat{\mu}_p$, $\mu_{p,N} \leq \hat{\mu}_p$. Therefore, eq. (\ref{eqn:minimized_power_mean_inequality}) holds.

Like the min-max type mass variables, $\mu_{p,N}$ has upper and lower bounds from its method of construction. 
\begin{eqnarray}
\label{eqn:upper_bound_minimized_power_mean}
\mu_{p,N} (m_\chi)
& \leq &
\hat{\mu}_p ( m_{A_1}, \cdots , m_{A_N} ) \; ,
\\
\label{eqn:lower_bound_minimized_power_mean}
\mu_{p,N}(\tilde{m}) \;\;
& \geq &
\hat{\mu}_p\left(m_{p_1} + \tilde{m}, \cdots ,m_{p_N} + \tilde{m} \right) \; .
\end{eqnarray}
The upper bound is defined in terms of the intermediate resonance masses only when the test masses of invisible particles are the correct invisible particle masses $m_\chi$. 
If a kinematic endpoint of $\mu_{p,N}$ distribution reaches this upper bound, i.e. \emph{saturated}, than we can interpret the kinematic endpoints into the true mass spectrum.   
In practice, saturation of events to the endpoint depends on physics process and hence a survey is needed.

Now we focus on a pair production of semi-invisibly decaying particles as figure \ref{fig:process_topology}, where $N=2$, then $\mu_{p,2}$ is defined as the following:
\begin{equation}
\mu_{p,2}
=
\min_{\substack{
	\mathbf{q}_{1}, \mathbf{q}_{2} \\ q_1^a + q_2^a = \slashed{p} {}_T^a
}}  
\hat{\mu}_p ( M_{A_1} , M_{A_2} ) \;.
\end{equation}
Unless the minimum is at a non-differentiable point, the invisible momentum solution satisfies the stationary condition,
\begin{equation}
\label{eqn:extremum_condition_longitudinal_raw}
\pdiff{}{q_i^z} \left. \hat{\mu}_p ( M_{A_1} , M_{A_2} ) \right|_{q_1^b + q_2^b = \slashed{p} {}_T^b} 
= 0 \; , 
\end{equation}
\begin{equation}
\label{eqn:extremum_condition_transverse_raw}
\pdiff{}{q_1^a} \left. \hat{\mu}_p ( M_{A_1} , M_{A_2} ) \right|_{q_1^b + q_2^b = \slashed{p} {}_T^b} 
= 0 \; , 
\end{equation}
For $p \geq 2$, these equations give the unique minimum of $\hat{\mu}_p ( M_{A_1} , M_{A_2} )$.
First, $(M_{A_i})^2$ is a convex function for invisible momenta because its Hessian matrix is positive semi-definite,
\begin{equation}
\label{eqn:Hessian_M_Square}
\det \frac{\partial^2 M_{A_i}^2}{ \partial q_i^a \partial q_i^b } 
= 
\frac{4 \tilde{m}^2 E_{p_i}^2 }{E_{q_i}^4} \geq 0 \;.
\end{equation}
Then $(M_{A_i})^{p}$ for $p \geq 2$ is a convex function for invisible momenta, and their sum $(M_{A_1})^{p} + (M_{A_2})^{p}$, which is $2 \hat{\mu}_p ( M_{A_1} , M_{A_2} )^p$, is also a convex function. 
Therefore, there is no local minimum, and the stationary condition can be used to find out the minimum of $\hat{\mu}_p ( M_{A_1} , M_{A_2} )$.
 
The minimization along the longitudinal direction is unconstrained, and thus two longitudinal component minimization can be performed independently. The condition is
\begin{equation}
\label{eqn:extremum_condition_longitudinal}
\hat{\mu}_p \cdot \frac{ M_{A_i}^{p-2} E_{p_i} ( \beta_{p_i}^z - \beta_{q_i}^z ) }{M_{A_1}^p + M_{A_2}^p } = 0 \; ,
\end{equation}
where $E_{\hat{p}}$ is a energy of a given momentum $p$ and $\beta_{p}^z$ is the longitudinal component of velocity of $p$. 
eq. (\ref{eqn:extremum_condition_longitudinal}) indicates that longitudinal velocities of $p_i$ and $q_i$ end up being identical after the minimization.
Then, each invariant masses $M_{A_i}$ turns into transverse masses with the solutions from the minimization of the longitudinal components of invisible momenta. 
Since $\mu_{p,2}$ is invariant under independent longitudinal boosts on resonance $A_1$ and $A_2$, 
we can boost each resonance and their daughter particles to a frame where all the longitudinal velocities of $p_i$ and $q_i$ vanishes, i.e. $\beta_{p_i}^z = \beta_{q_i}^z = 0$. 
Since all the longitudinal components on eq. (\ref{eqn:extremum_condition_transverse_raw}) vanishes in this frame, using this frame is convenient for further discussion.

The minimization along the transverse direction is constrained by the missing transverse momentum, and hence the extremum condition, eq. (\ref{eqn:extremum_condition_transverse_raw}), gives a nontrivial relation. 
As long as $M_{A_i}$ are differentiable at the minimum, the condition can be written as a mass ratio $M_{A_1} / M_{A_2}$, 
\begin{equation}
\label{eqn:extremum_condition_ratio}
\left( \frac{M_{A_1}}{M_{A_2}} \right)^{p-2} = \frac{ E_{p_2} ( \beta_{p_2}^a - \beta_{q_2}^a ) }{ E_{p_1} ( \beta_{p_1}^a - \beta_{q_1}^a ) }  \;,
\end{equation}
where $E_{p}$ is an energy of a given momentum $p$ and $\beta_{p}^{a}$ is a transverse component of velocity of $p$.

More specifically, for $p=2$ and $\tilde{m} = 0$, eq. (\ref{eqn:extremum_condition_ratio}) is a system of simple polynomial equations, and thus, $\mu_{2,2}(0)$ has an analytic solution. 
Applying the method of Lagrange multipliers on eq. (\ref{eqn:extremum_condition_transverse_raw}), we can find out two non-trivial solutions of missing momenta, which are obtained for $\bar{D} \geq 0$,
\begin{equation*}
\hat{q}_1^a = 
\frac{1}{E_{p_1}} \left[ \frac{\Delta_m + 2 p_1 \cdot \Delta}{2 \Delta \cdot \Delta} \Delta^a \pm \frac{\sqrt{\bar{D}}}{2} \left( \sigma^a - \frac{\sigma \cdot \Delta}{\Delta \cdot \Delta} \Delta^a \right) \right] ,
\end{equation*}
\begin{equation}
\hat{q}_2^a = 
\frac{1}{E_{p_2}}
\left[ \frac{\Delta_m + 2 p_2 \cdot \Delta}{2 \Delta \cdot \Delta} \Delta^a \pm \frac{\sqrt{\bar{D}}}{2} \left( \sigma^a - \frac{\sigma \cdot \Delta}{\Delta \cdot \Delta} \Delta^a \right)
\right] ,
\end{equation}
where
\begin{eqnarray*}
\hat{q}_i^a = \frac{q_i^a}{|q_i|} = \left. \beta^a_{q_i} \right|_{\tilde{m} = 0} & \;,\quad & |q_i| = \sqrt{(q_i^1)^2 + (q_i^2)^2} \;, \\
\sigma^a = p_1^a + p_2^a & \;,\quad & \sigma_m = m_{p_1}^2 + m_{p_2}^2 \;, \\
\Delta^a = p_1^a - p_2^a & \;,\quad & \Delta_m = m_{p_1}^2 - m_{p_2}^2 \;, 
\end{eqnarray*}
\begin{eqnarray*}
\bar{D} & = & 1 - \frac{4c}{a} \;, \\
a & = & \sigma\cdot \sigma - \frac{(\sigma \cdot \Delta)^2}{\Delta \cdot \Delta} \;, \\
c & = & \frac{\Delta_m^2}{4 \Delta \cdot \Delta} + \frac{\Delta_m}{2 \Delta \cdot \Delta} \sigma \cdot \Delta - \frac{\sigma_m}{2} \;.
\end{eqnarray*}
and $u \cdot v = u^a v_a$. The magnitude $|q_i|$ of transverse-projected invisible momenta $q_i^a$ can be obtained from the transverse momentum conservation, $q_1^a + q_2^a = \slashed{p} {}_T^a$. Decomposing $q_i^a$'s into magnitudes and unit vectors, i.e. $q_i^a = |q_i| \hat{q}_i^a$, the solution is
\begin{equation}
\left(
\begin{matrix}
|q_1|  \\
|q_2|
\end{matrix}
\right)
=
\frac{1}{\hat{q}_1 \times \hat{q}_2}
\left(
\begin{matrix}
\slashed{p}_T \times \hat{q}_2  \\
- \slashed{p}_T \times \hat{q}_1
\end{matrix}
\right) \;,
\end{equation}
where $u \times v = \epsilon_{ab} u^a v^b$. If $\bar{D} < 0$ or $|q_i| < 0$, then the solution is not physical. If non-trivial solutions do not exist, or both are not physical, then the minimum is located at one of two non-differentiable points where $q_1^a = 0$ or $q_2^a = 0$.
For other cases of $p$ with arbitrary trial masses $\tilde{m}$ of invisible particles, eq. (\ref{eqn:extremum_condition_ratio}) is a complicated radical equation. Analytically solving the minimization is not much illuminating at this moment, we will solve the minimization numerically instead.

In the limit of $p$ to $\infty$, by definition, the $\mu_{p,2}$ converges to $M_2$ variable with minimal constraints, which is identical to the Cambridge-$M_{T2}$ variable
\begin{equation}
\label{eqn:definition_M2}
\mu_{\infty,2} = \lim_{p \rightarrow \infty} \mu_{p,2} = \min_{\substack{
	\mathbf{q}_{1}, \mathbf{q}_{2} \\ 
	q_1^a + q_2^a = \slashed{p} {}_T^a
}}   \max ( M_{A_1} , M_{A_2} ) \;.
\end{equation}
In this limit, the extremum condition in eq. (\ref{eqn:extremum_condition_ratio}) converges to the two types of solutions, the balanced and unbalanced solutions \cite{Barr:2003rg, Cho:2007dh, Lester:2007fq}. If the minimum is developed where the differences of velocity $\beta_{p_i}^{a} - \beta_{q_i}^{a}$ are non-zero, the limit of eq. (\ref{eqn:extremum_condition_ratio}) for $p \rightarrow \infty$ requires that the reconstructed mass ratio $M_{A_1}/M_{A_2}$ should converge to 1. This mass ratio corresponds to the balanced solution of $\mu_{\infty,2}$, i.e. $M_{T2}$. If one of the differences of velocity approaches to zero at the minimum, the mass ratio does not need to converge to 1 anymore. This momenta configuration is located at a minimum of corresponding $M_{A_i}$, and it is the unbalanced solution of $\mu_{\infty,2}$. 
\\

%
\section{Separation of the Kinematic Endpoints}
%
To see the behavior of kinematic endpoint of $\mu_{p,2}$, we generated event samples by a phase-space Monte Carlo program for two-body and three-body decays whose visible particles are assumed to be massless. 
For understanding behavior of a kinematic endpoint, pure phase-space analysis is enough because matrix elements, which includes effects of spin-correlation and coupling structures, only alter a shape of given distribution. 
These effects could affect accuracy of measuring the kinematic endpoint, but this systematic error is less relevant in the case of the transverse mass variables, which includes $M_{T}$ and $M_{T2}$.
Their shape near the endpoint is dominated by phase-space effects \cite{Giudice:2011ib}, such as the number of missing particles and masses of missing particles.
Furthermore, the whole shape of $M_{T2}$ distribution also does not depend on spins of each intermediate and final state particles much \cite{Edelhauser:2012xb}.
Since $\mu_{p,2}$ also relies on transverse mass variables, we expect that pure phase-space based survey would be enough for understanding the kinematic endpoint behavior.

To check dependence of endpoint saturation on a total center-of-mass energy $\sqrt{\hat{s}}$ of colliding partons, we further assume that $\sqrt{\hat{s}}$ is fixed for all events. 
Upstream momentum including initial state radiation is not considered in this analysis. 
$\mu_{\infty,2}$ is evaluated by a bisection-based method \cite{Cheng:2008hk, Lester:2014yga}. 
The other $\mu_{p,2}$'s are calculated by a variable metric method equipped with additional derivative information using {\sc Minuit2} \cite{James:1975dr}. 
We further assume that all the visible particles are correctly assigned to their mother resonance, and the true invisible particle masses $m_{\chi}$ are supplied to the trial invisible particle masses $\tilde{m}$.

\begin{figure}[t]
\centering
\begin{tabular}{cc}
\includegraphics[width=5.5cm]{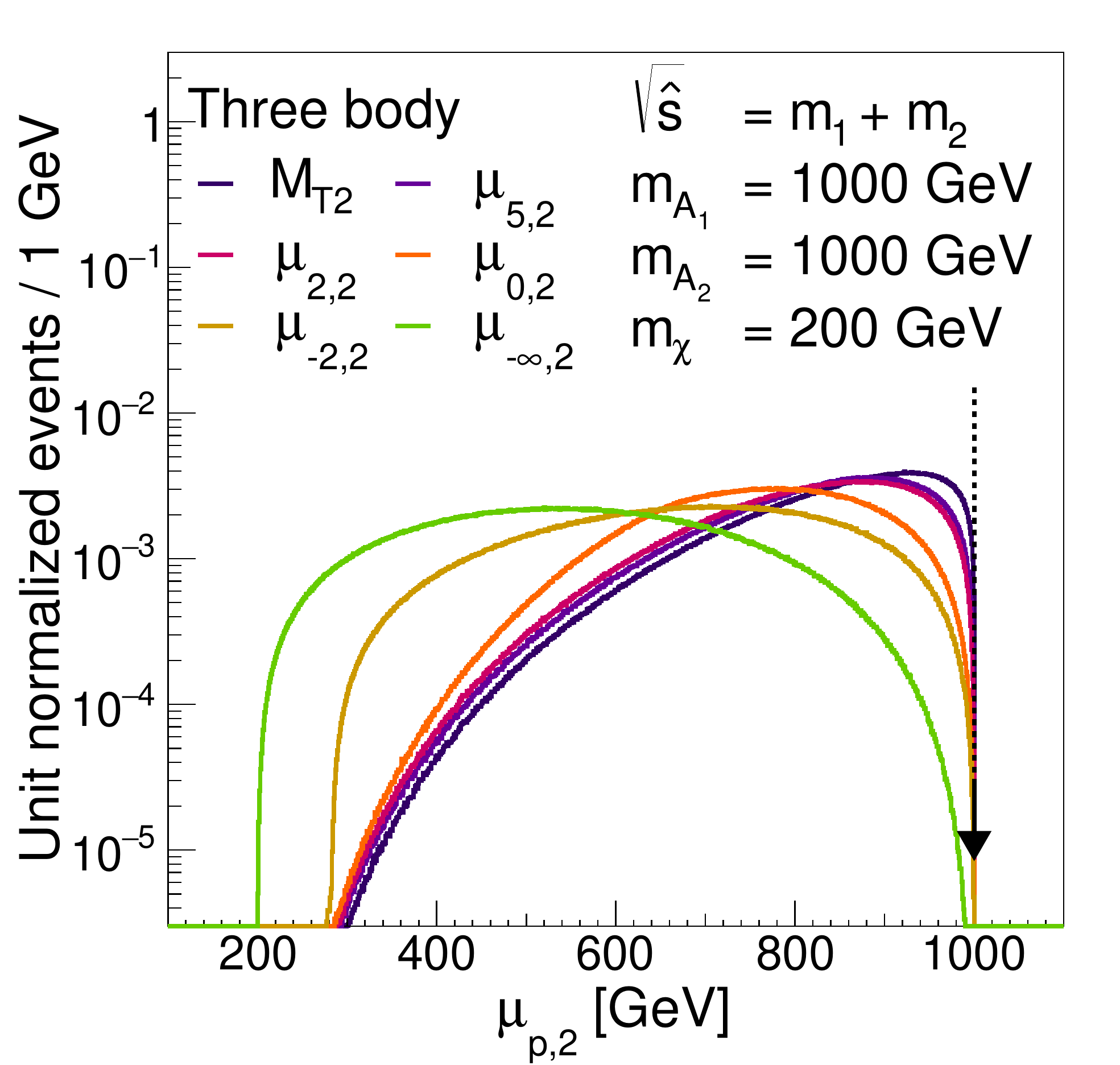} &
\includegraphics[width=5.5cm]{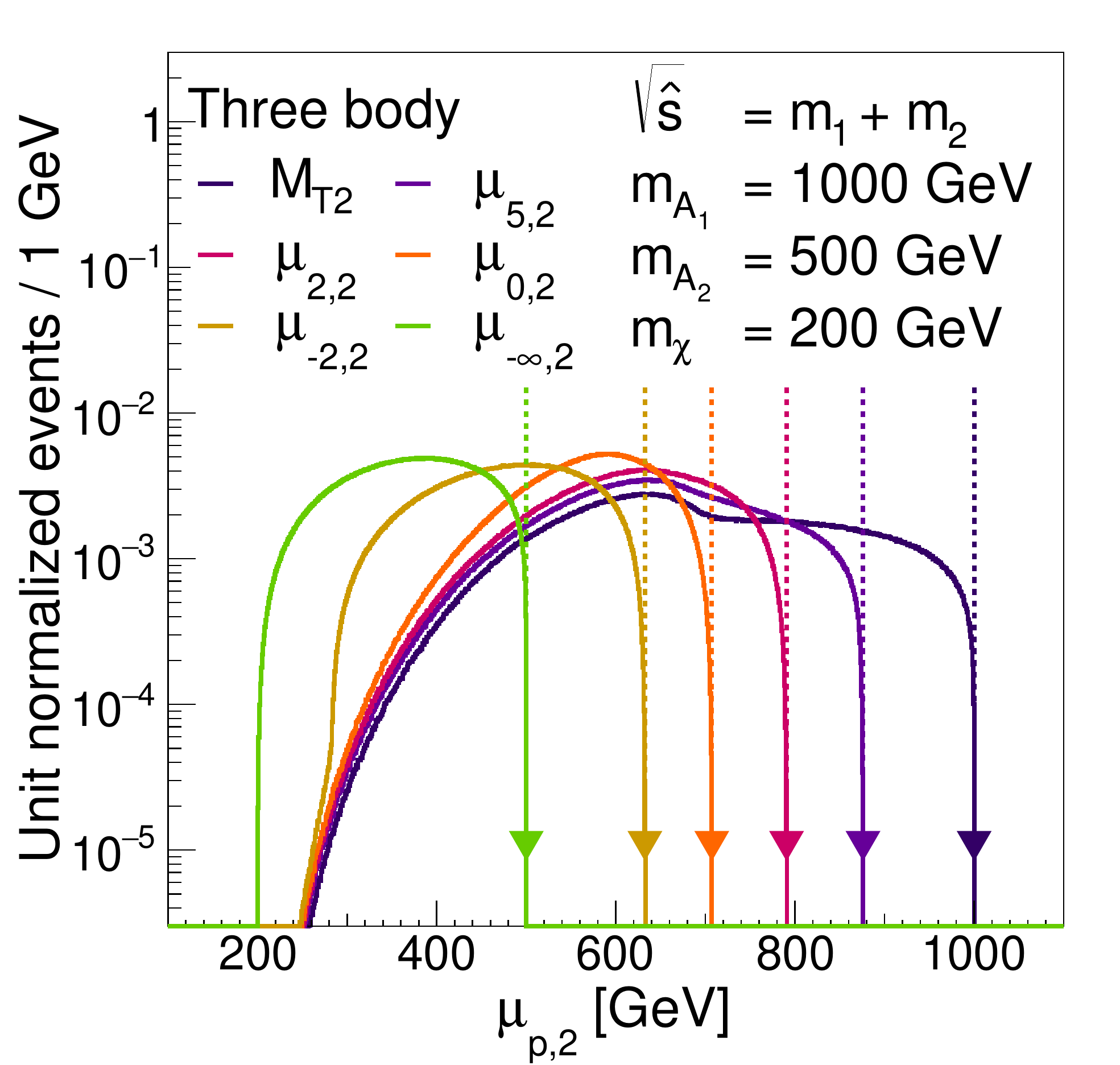} 
\end{tabular}
\caption{\label{fig:sample_distribution_threebody_threshold}$\mu_{p,2}$ distribution in a three-body decay for a threshold production. The left histogram represents a symmetric mass spectrum, and the right histogram represents an asymmetric mass spectrum. Arrows show the locations of the upper bounds for each distribution.}
\end{figure}
\begin{figure}[t]
\centering
\begin{tabular}{cc}
\includegraphics[width=5.5cm,keepaspectratio=true]{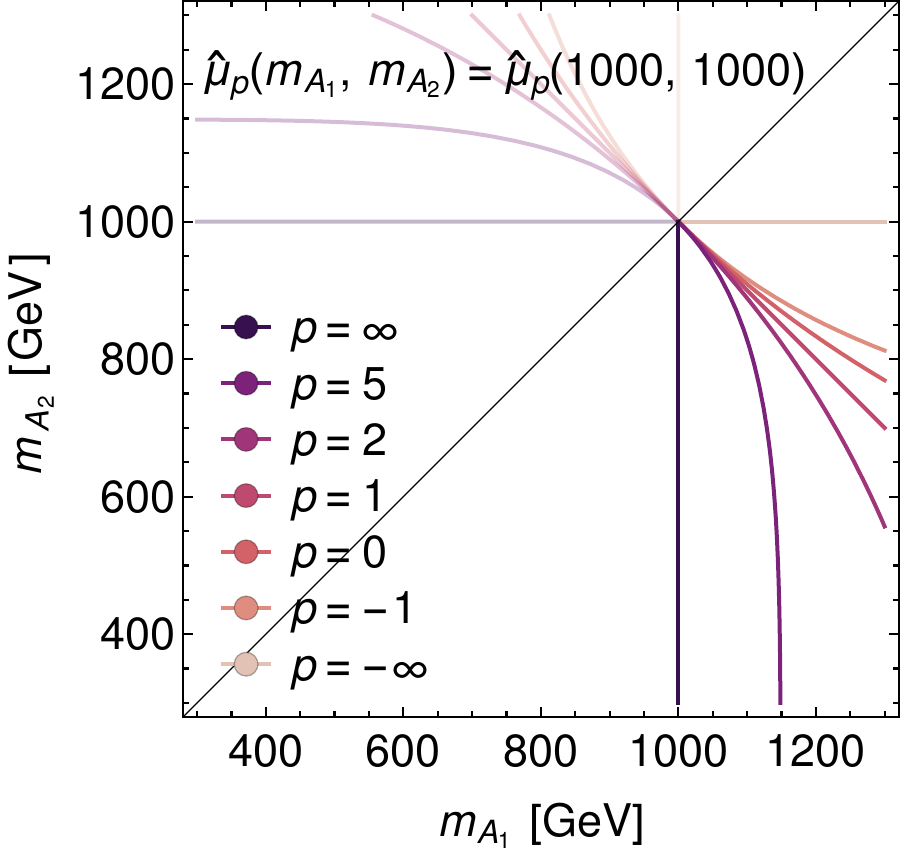} &
\includegraphics[width=5.5cm,keepaspectratio=true]{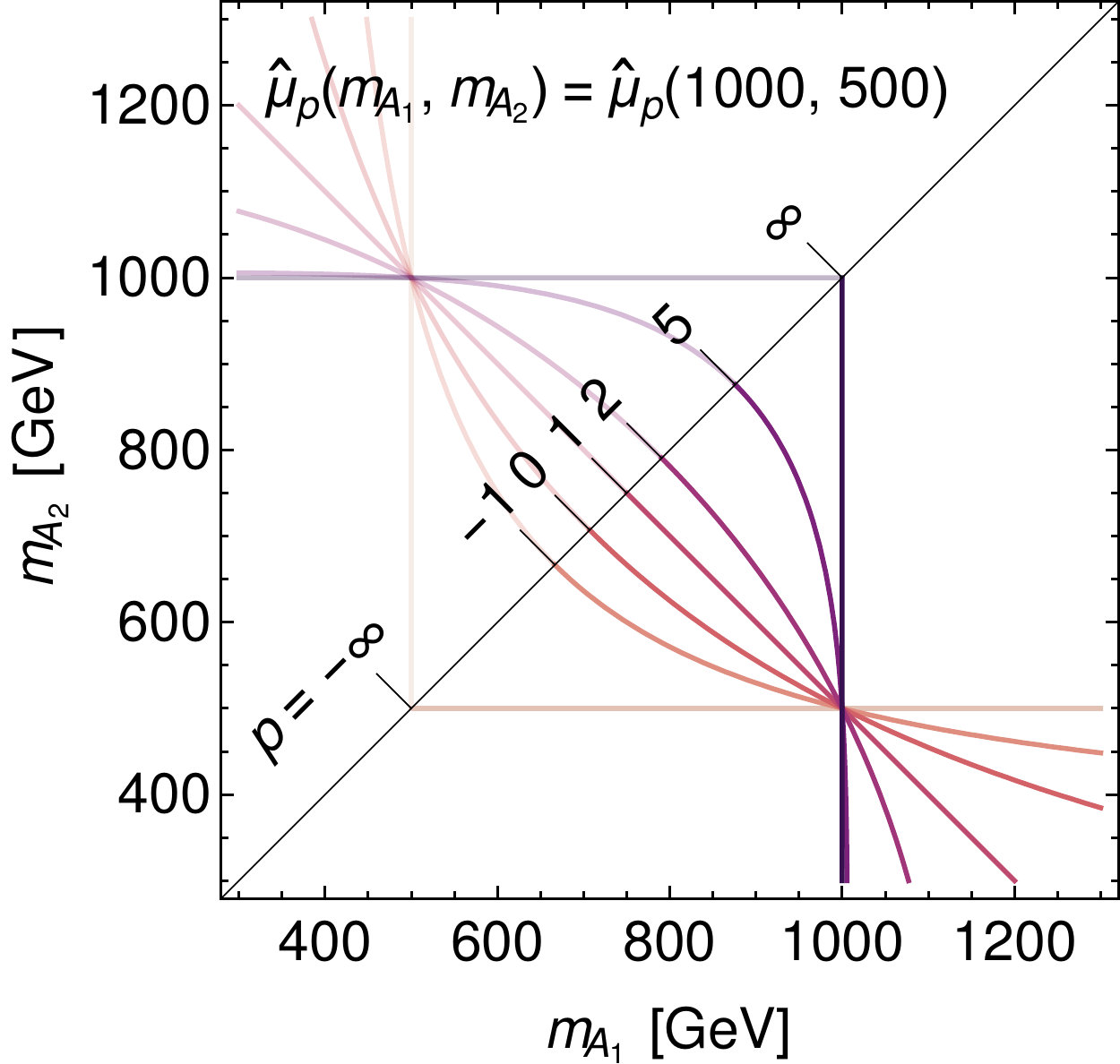} 
\end{tabular}
\caption{\label{fig:mass_fit_threebody_threshold}Contours of $\hat{\mu}_p(m_{A_1},m_{A_2})$ on an $(m_{A_1},m_{A_2})$ plane such that 
$\hat{\mu}_p$ takes the value of the corresponding kinematic endpoint of $\mu_{p,2}$. The left contours represent a symmetric mass spectrum, and the right contours represent an asymmetric mass spectrum. Because these contours over-constrain mass spectra, their intersection eventually crosses the true mass spectrum.
}
\end{figure}

Let us consider a threshold production first. In this case, both resonances are in their rest frame and hence the transverse missing energy has limited chance to have a larger value. Since $\mu_{\infty,2}$ tends to have larger value for large missing transverse energy, this is a minimal checkpoint to test saturation of the kinematic endpoint. 

Three-body decay is an ideal case for mass measurement because most of $\mu_{p,2}$ distributions have well-saturated kinematic endpoints for both symmetric and asymmetric mass spectrum if correct trial masses $m_\chi$ of invisible particles are supplied. 
In a three-body decay, $m_{p_i}$ distribution is known to have a kinematic endpoint $m_{p_i}^{\max} = m_{A_i} - m_\chi$. 
Because $\mu_{p,2}$ is bounded below by eq. (\ref{eqn:lower_bound_minimized_power_mean}), events near the kinematic endpoint of $m_{p_i}$ have $\mu_{p,2}$ bounded below by $\hat{\mu}_p ( m_{A_1} - m_\chi + \tilde{m} , m_{A_2} - m_\chi + \tilde{m} )$.
If the trial invisible particle mass $\tilde{m}$ is the true $m_{\chi}$, then the $\mu_{p,2}$ is bounded below by $\hat{\mu}_p (m_{A_1}, m_{A_2})$. 
Therefore, events near the kinematic endpoint of $m_{p_i}$ distribution have $\mu_{p,2}$ value near $\hat{\mu}_p (m_{A_1}, m_{A_2})$, and the kinematic endpoint of $\mu_{p,2}$ can be saturated as figure \ref{fig:sample_distribution_threebody_threshold}.
If these kinematic endpoints are not hidden behind backgrounds, then the kinematic endpoints are extractable from $\mu_{p,2}$ distributions. 
 
Fitting the extracted endpoints by the power mean $\hat{\mu}_p (m_{A_1}, m_{A_2})$ will reveal the true mass spectrum like figure \ref{fig:mass_fit_threebody_threshold}. 
Each fitted kinematic endpoints constrain mass spectra independently, and the constrained line eventually cross at a single point because the kinematic endpoints over-constrain the true mass spectrum. In a realistic case, the fitting always comes with errors and hence the constrained lines become constrained bands. The intersection of these bands constrains the true mass spectrum then.

\begin{figure}[t]
\centering
\begin{tabular}{cc}
\includegraphics[width=5.5cm]{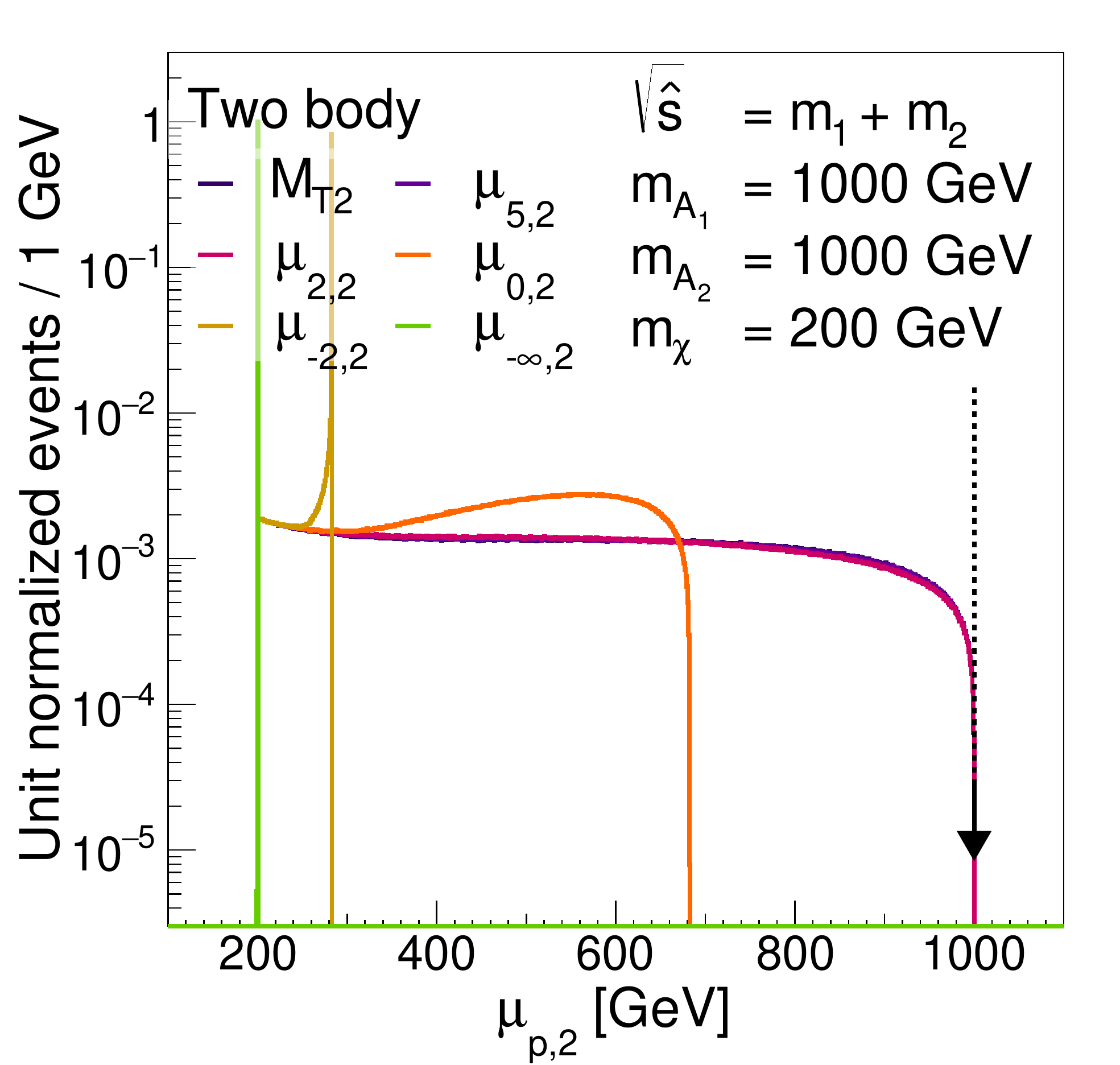} &
\includegraphics[width=5.5cm]{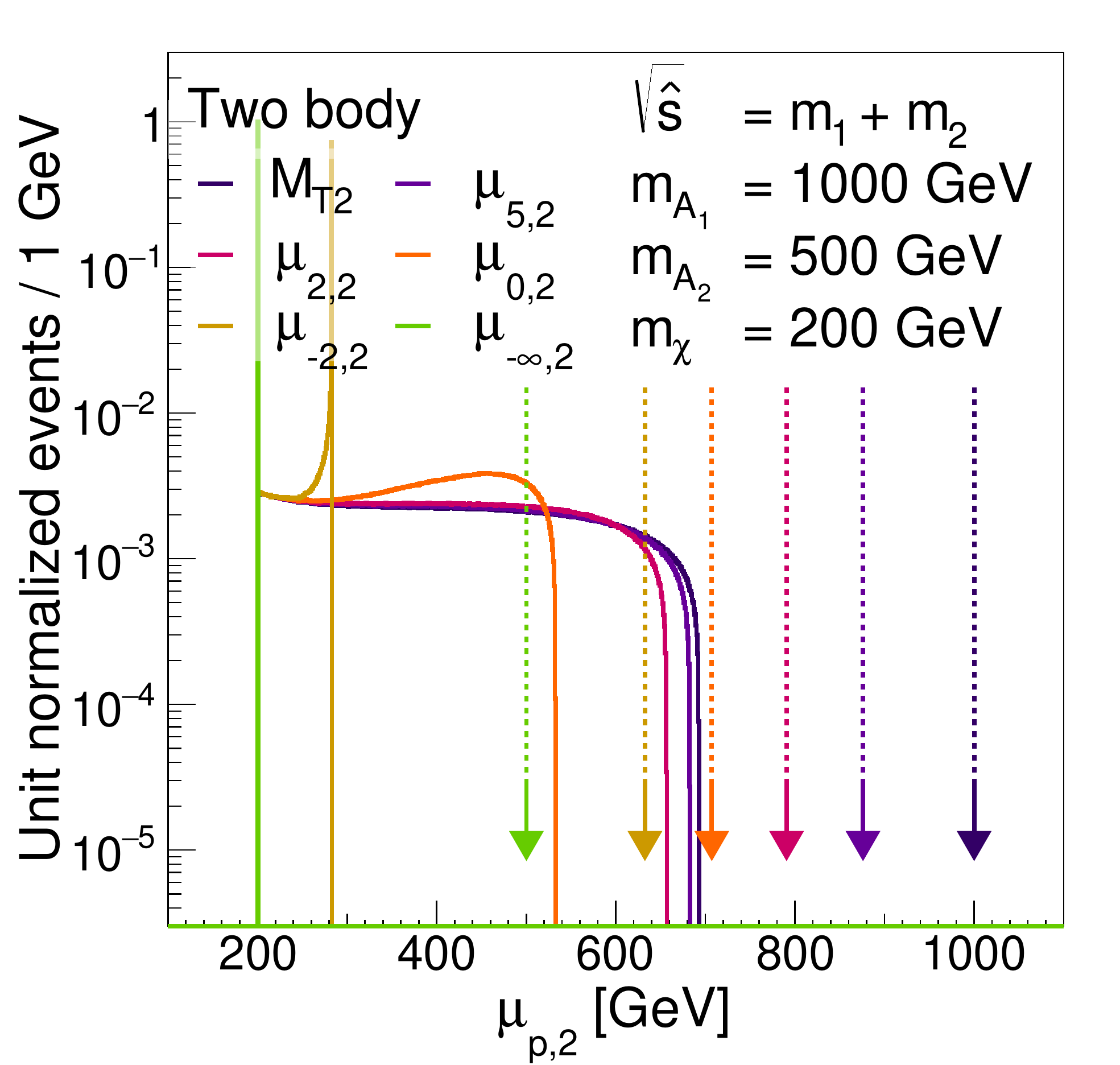} \\
\includegraphics[width=5.5cm]{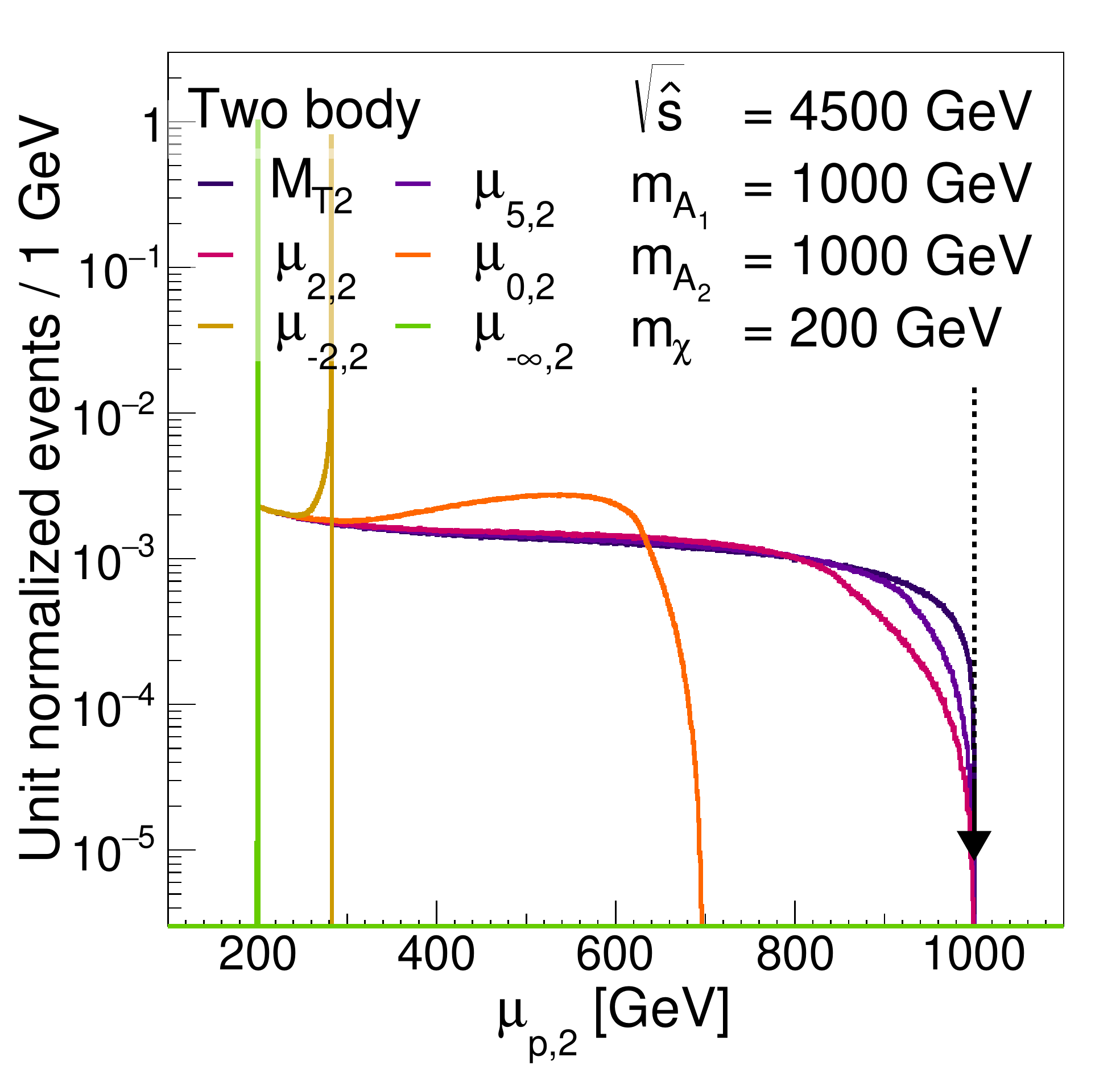} &
\includegraphics[width=5.5cm]{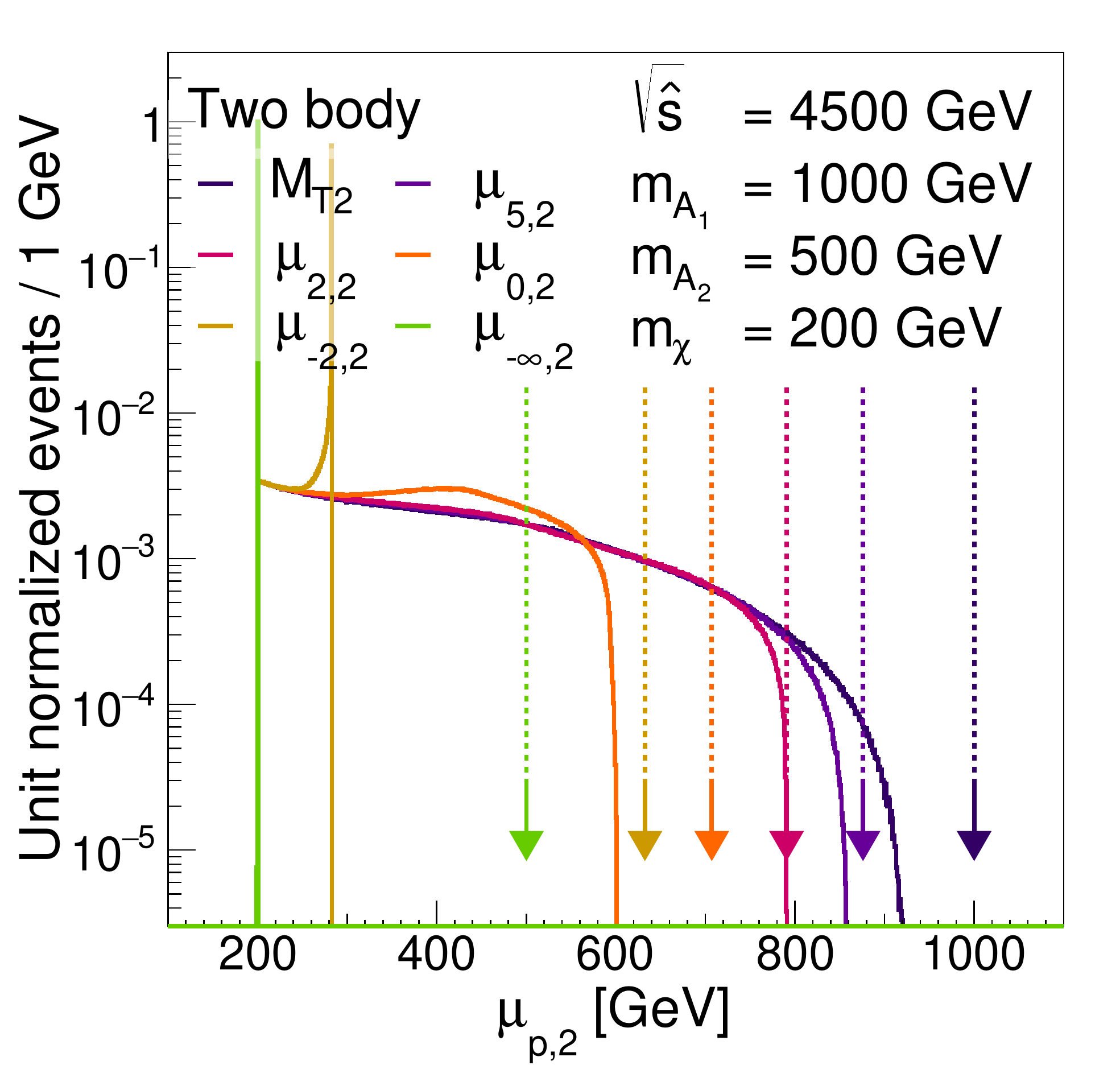} 
\end{tabular}
\caption{\label{fig:sample_distribution_twobody_threshold}$\mu_{p,2}$ distribution in a two-body decay. for a threshold production. The left histograms represent a symmetric mass spectrum and the right histograms represent an asymmetric mass spectrum. The upper histogram are for a threshold production, and the lower histograms for an energetic production. Arrows show the locations of the upper bounds for each distribution.}
\end{figure}

Two-body decay case is harder to have a well-saturated endpoint. Because there is no assist from $m_{p_i}$ distributions, only some of the $\mu_{p,2}$ distributions saturates their upper bound as figure \ref{fig:sample_distribution_twobody_threshold}. 
For a threshold production, the kinematic endpoint of $\mu_{\infty,2}$ distribution is saturated only when the upper bound only when the true mass spectrum is symmetric.
If the true mass spectrum is asymmetric, $\mu_{\infty,2}$ distribution does not reach the upper bound.
$\mu_{p,2}$ also have the same problem because $\mu_{p,2}$ is increasing function of $p$ as eq. (\ref{eqn:minimized_power_mean_inequality}). 
If the kinematic endpoint $\mu_{\infty,2}^{\mathrm{max}}$ of $\mu_{\infty,2}$ distribution is smaller than the upper bound of $\mu_{p,2}$, then the kinematic endpoint of $\mu_{p,2}$ is not saturated because any of $\mu_{p,2}$ cannot exceed $\mu_{\infty,2}^{\mathrm{max}}$.

Although the kinematic endpoint is not saturated in an asymmetric mass spectrum, there is a difference between the endpoints of the distributions in both mass spectrum. For $p \gtrsim 2$, the kinematic endpoint of $\mu_{p,2}$ distribution matches in symmetric mass spectrum while it is not in an asymmetric mass spectrum. Therefore, we can utilize this separated kinematic endpoints to resolve general asymmetric mass spectrum.

\begin{figure}[t]
\begin{center}
\includegraphics[width=5.5cm]{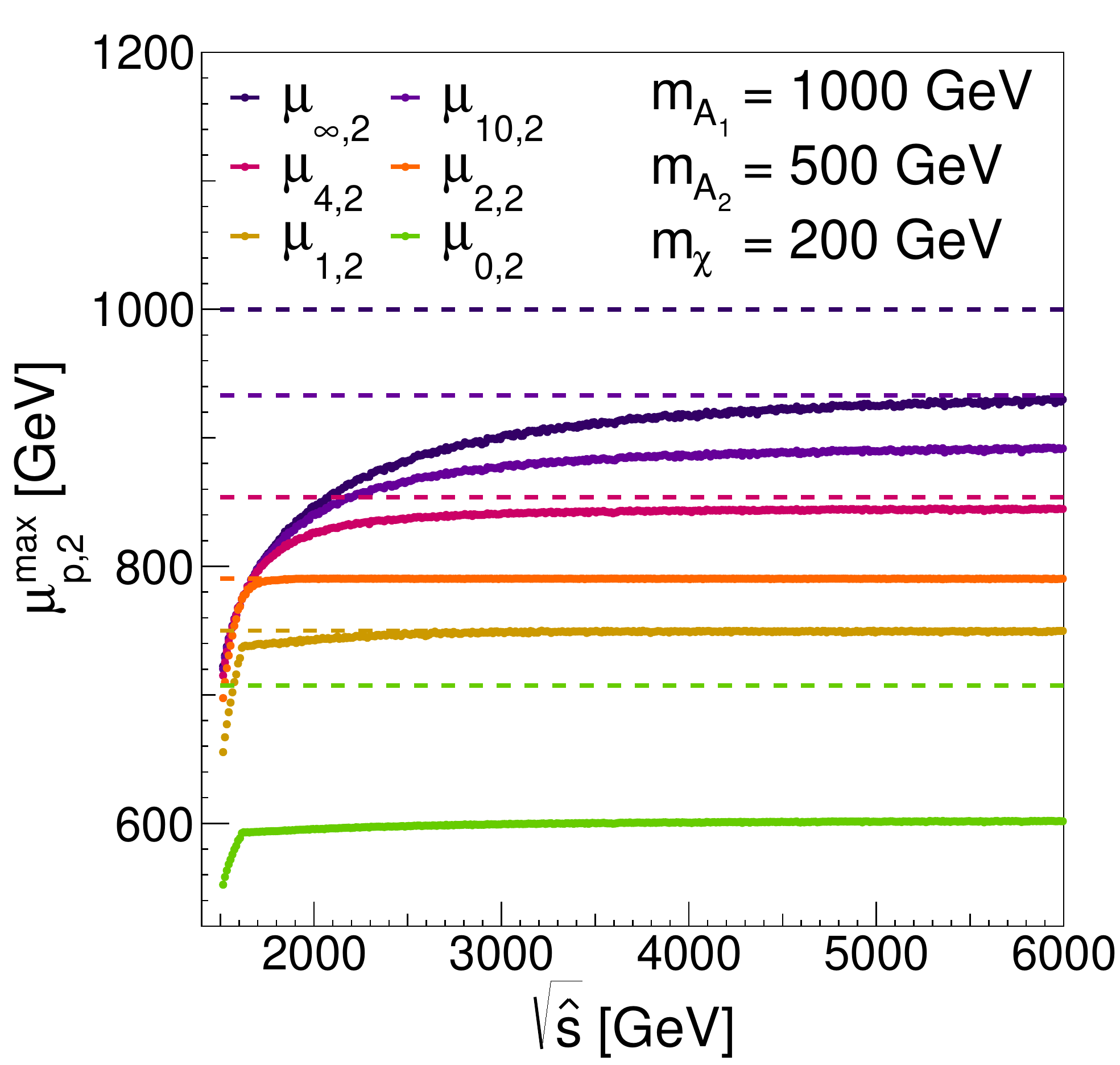} 
\end{center}
\caption{\label{fig:endpoint_sqrts_dependence}Total center-of-mass energy $\sqrt{\hat{s}}$ of colliding partons dependence of the maximum point $\mu_{p,2}^{\max}$ of $\mu_{p,2}$ distribution in a two-body decay. The dashed line represents the upper bound of the corresponding $\mu_{p,2}$. For $m_{A_{1}} = 1000$ GeV and $m_{A_{2}} = 500$ GeV, as soon as $M_{T2}$ exceed the upper bound of $\mu_{2,2}$, $\mu_{2,2}$ have a saturated kinematic endpoint. }
\end{figure}

\begin{figure}[t]
\begin{center}
\includegraphics[width=5.5cm,keepaspectratio=true]{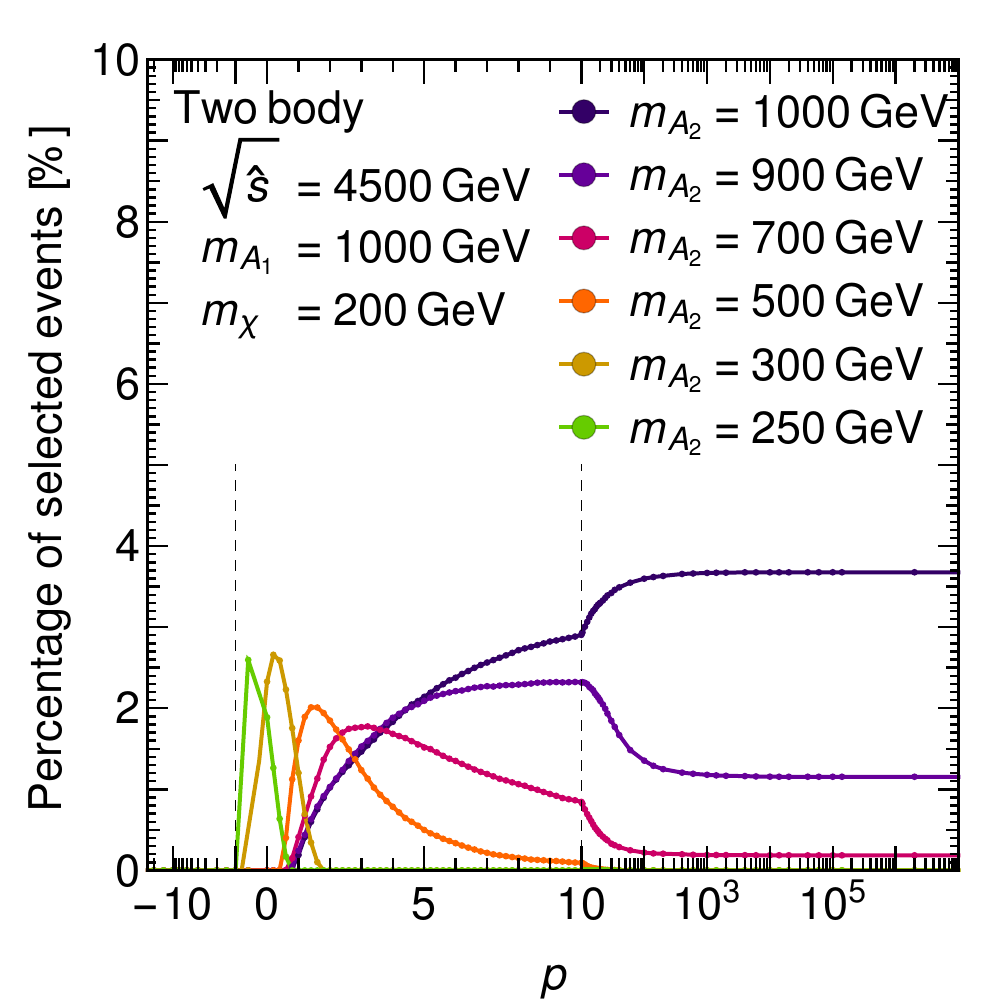} 
\end{center}
\caption{\label{fig:endpoint_power_dependence}
Percentage of events having $\mu_{p,2}$ at least 90\% of the upper bound value of corresponding $\mu_{p,2}$ in a two-body decay is drawn for various mass spectra. 
For $p$ between -1 and 10, the domain is drawn in linear scale, and other domains are drawn in log scale. 
In general, $\mu_{p,2}$ which shows best endpoint saturation is mass spectrum dependent.}
\end{figure}

This feature is enhanced when we consider a pair production at higher energy. 
For an asymmetric pair production at some large $\sqrt{\hat{s}}$, figure \ref{fig:sample_distribution_twobody_threshold} shows some $\mu_{p,2}$ saturates its upper bound. 
In particular, for $m_{A_1} = 1000 \; \mathrm{GeV}$, $m_{A_2} = 500 \; \mathrm{GeV}$ and $m_{\chi} = 200 \; \mathrm{GeV}$, $\mu_{2,2}$ shows a good saturation. 
To check $\sqrt{\hat{s}}$ dependence of the saturation of the kinematic endpoints for this mass spectrum, we draw $\mu_{p,2}^{\max}$ versus $\sqrt{\hat{s}}$ in figure \ref{fig:endpoint_sqrts_dependence}.
We can observe that there exist some $\mu_{p,2}$ such that its kinematic endpoint is saturated as soon as $\mu_{\infty,2}^{\mathrm{max}}$ exceed the upper bound of $\mu_{p,2}$ distribution.
Once some $\mu_{p,2}$ has the saturated kinematic endpoint, the kinematic endpoint does not change when we consider higher $\sqrt{\hat{s}}$ while the kinematic endpoint of other $\mu_{p,2}$ distribution has more chance to have larger $\mu_{p,2}$ value. 
Therefore, the separated kinematic endpoint is a generic feature of an asymmetric mass spectrum.

\begin{figure}[t]
\centering
\begin{tabular}{cc}
\includegraphics[width=5.5cm]{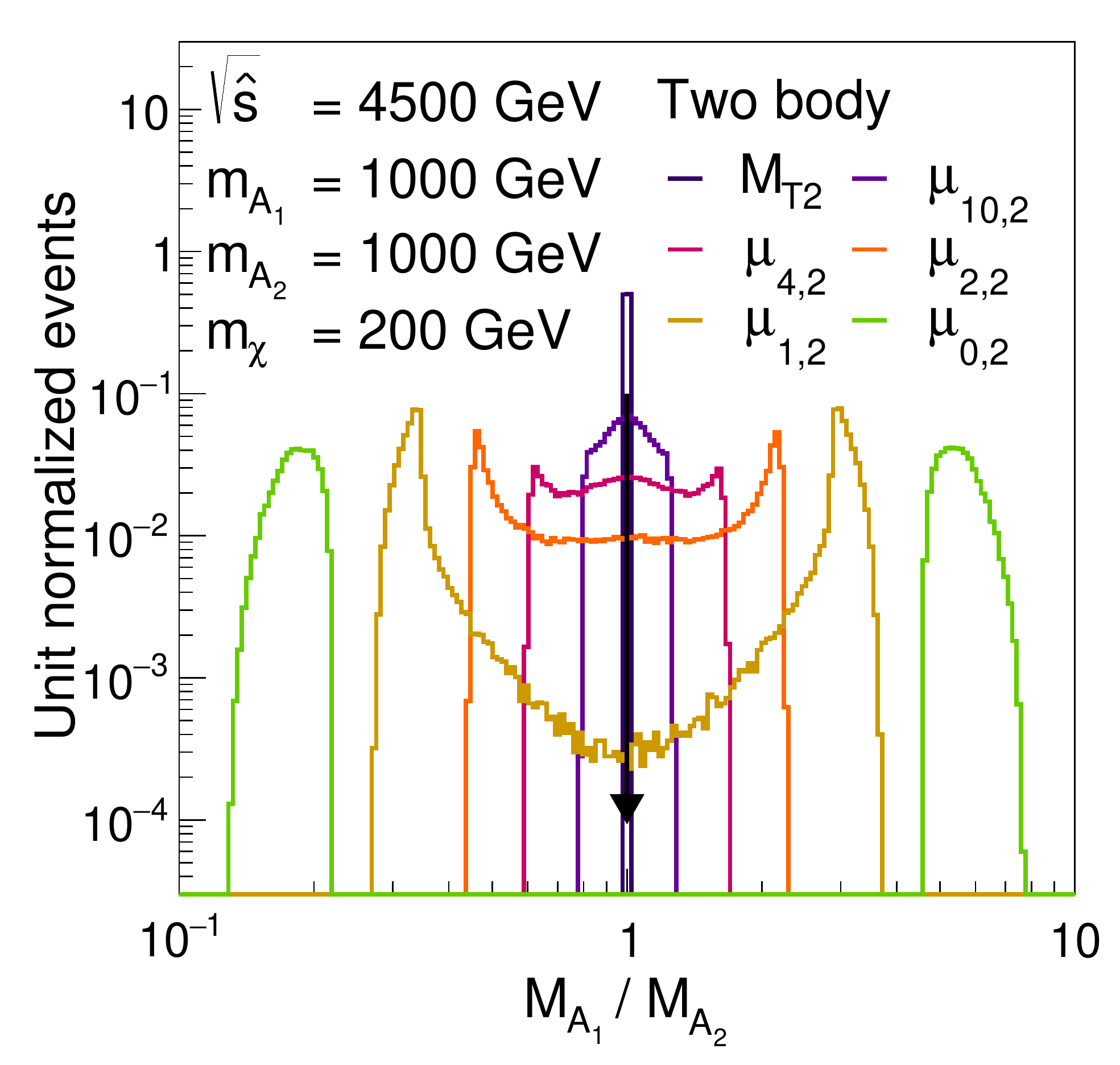} &
\includegraphics[width=5.5cm]{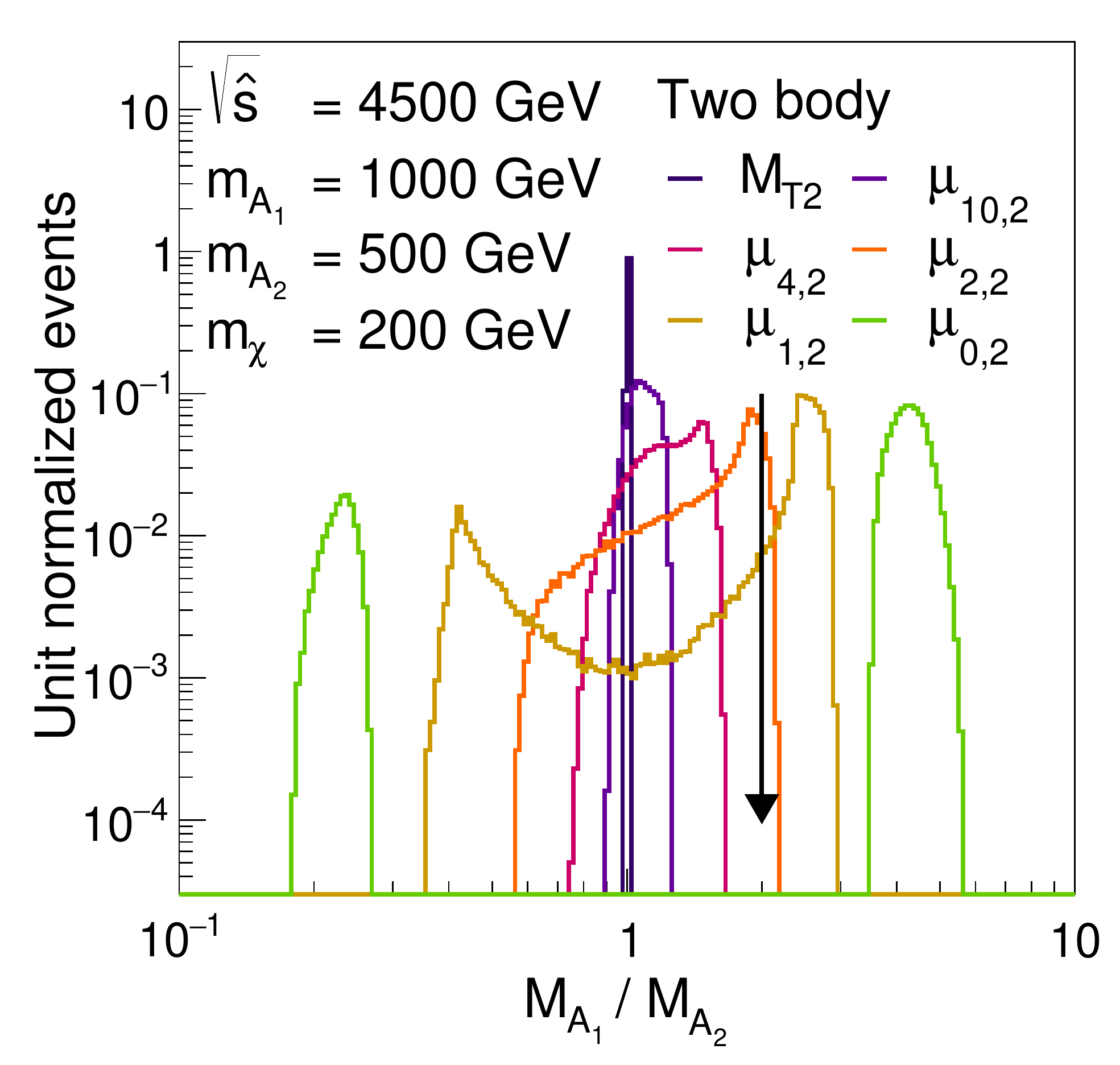}
\end{tabular}
\caption{\label{fig:endpoint_mass_ratio}The ratio of reconstructed masses for near endpoint events in a two-body decay. We select events having the top 10\% mass value. 
Arrows on each plot point $\frac{m_1}{m_2}$. 
By comparing with figure \ref{fig:endpoint_power_dependence}, the number of events near the kinematic endpoint of $\mu_{p,2}$ distribution having correctly reconstructed mass ratio and saturation of the kinematic endpoint of $\mu_{p,2}$ have a positive correlation.}
\end{figure}

A particular power $p$ of $\mu_{p,2}$ distribution which saturates its upper bound is mass spectrum dependent. 
Figure \ref{fig:endpoint_power_dependence} shows there are specific $\mu_{p,2}$'s having saturated kinematic endpoints. 
The reason the power $p$ changes is that only some $p$ supports mass ratio in eq. (\ref{eqn:extremum_condition_ratio}) compatible to the true mass spectrum for near endpoint events. 
Suppose $\mu_{p,2}$ can saturate the upper bound, i.e. there exist some events whose $\mu_{p,2}$ solution satisfy
\begin{equation}
\hat{\mu}_p ( M_{A_1} , M_{A_2} ) = \hat{\mu}_p ( m_{A_1} , m_{A_2} ) 
\end{equation}
Let $M_{A_1} / M_{A_2} = c \, m_{A_1} / m_{A_2}$ where $c$ is a positive constant. Then above equation can be written as a mass ratio between the trial resonance mass $M_{A_i}$ and the true resonance mass $m_{A_i}$. 
\begin{equation}
\frac{M_{A_1}}{ m_{A_1}} = \frac{ \hat{\mu}_p ( m_{A_1} , m_{A_2} ) }{ \hat{\mu}_p (  m_{A_1} , c^{-1}m_{A_2} ) }
, \quad
\frac{M_{A_2}}{ m_{A_2}} = \frac{ \hat{\mu}_p ( m_{A_1} , m_{A_2} ) }{ \hat{\mu}_p ( c \, m_{A_1} , m_{A_2} ) }
\end{equation}
Unless $c=1$, $M_{A_1} \neq m_{A_1}$ and $M_{A_2} \neq m_{A_2}$, and hence invisible momenta $q_i$ of this $\mu_{p,2}$ solution are different to the true invisible momenta. 
Those two $q_i$ configurations are two distinct minima of $\hat{\mu}_p ( M_{A_1} , M_{A_2} )$ then. 
Recall that $\hat{\mu}_p ( M_{A_1} , M_{A_2} )^p$ is a convex function for $p \geq 2$ as eq. \ref{eqn:Hessian_M_Square}, and it is a strictly convex function when $\tilde{m} \neq 0$. 
If $m_{\chi} \neq 0$, then having multiple minima is a contradiction because a strictly convex function does not admit multiple minima.
Let $m_\chi = 0$. Then a determinant of Hessian matrix of $(M_{A_i})^p$ is
\begin{equation}
\label{eqn:Hessian_M_p_massless}
\det \frac{\partial^2 ( M_{A_i} )^{p}}{ \partial q_i^a \partial q_i^b } 
=
\frac{8 p^2 (p-2)(M_{A_i})^{2p-6} E_{p_i}}{E_{q_i}} (E_{p_i} - (p_i \cdot \hat{q}_i))^2 \;.
\end{equation}
If $p > 2$ and one of the $m_{p_i}$ is nonzero, then the determinant of Hessian matrix of $(M_{A_i})^2$ is positive definite, $\hat{\mu}_p ( M_{A_1} , M_{A_2} )^p$ is strictly convex function, and it is a contradiction. If both $m_{p_i} = 0$, the determinant can be zero when $p_i$ and $q_i$ are heading the same direction, i.e. $\hat{p}_i = \hat{q}_i$. 
If this direction of $M_{A_1}$ and $M_{A_2}$ overlap, then $\hat{\mu}_p ( M_{A_1} , M_{A_2} )^p$ is just convex on the intersection and it can admit multiple minima. 
However, $\hat{\mu}_p ( M_{A_1} , M_{A_2} )^p = 0$ along this line and hence such events cannot locate near the upper bound. 
For $p = 2$, the determinant is zero, and hence, $\hat{\mu}_2 ( M_{A_1} , M_{A_2} )^2$ is not a strictly convex function.
The previous argument is not applicable in this case, but we can argue the existence of multiple solutions easily because $M_{A_i}^2$ is a linear function in $|q_i|$.
\begin{equation}
M_{A_i}^2 = m_{p_i}^2 + |q_i| ( E_{p_i} - p_i \cdot \hat{q_i} ) \;.
\end{equation}  
When $m_{p_i} = 0$, there is flat direction, $\hat{p}_i = \hat{q}_i$. Again, multiple minima can be found if this direction is aligned, but such events cannot locate near the kinematic endpoint.
In this case, there is another type of multiple minima located between a line between $q_1^a = 0$ and $q_1^a = \slashed{p} {}_T^a$ on $q_i$ space. 
If slopes of $M_{A_1}^2$ and $M_{A_2}^2$ on the line exactly cancel out, then $\hat{\mu}_2 ( M_{A_1} , M_{A_2} )^2$ is flat along this line, and this line can be multiple minima. 
The condition can be written by 
\begin{equation}
(p_1 + p_2) \cdot \hat{\slashed{p}}_T = E_{p_1} - E_{p_2}
\end{equation}
Events satisfying above condition are the only cases where $\hat{\mu}_2 ( M_{A_1} , M_{A_2} )^2$ admit multiple minima. 

Therefore, at least for $p > 2$ and $p = 2$ except some special situations, we can conclude if the mass ratio in eq. (\ref{eqn:extremum_condition_ratio}) does not support the true mass spectrum, then $\mu_{p,2}$ cannot saturate the upper bound.
We draw reconstructed mass ratio of events near the kinematic endpoint of $\mu_{p,2}$ distribution in figure \ref{fig:endpoint_mass_ratio}.
By comparing figure \ref{fig:endpoint_power_dependence} and figure \ref{fig:endpoint_mass_ratio}, we can see a correlation between the saturation of the kinematic endpoint of $\mu_{p,2}$ and number of events near the kinematic endpoint of $\mu_{p,2}$ distribution, which have correctly reconstructed mass ratio.

Furthermore, the strict convexity of $\hat{\mu}_p ( M_{A_1} , M_{A_2} )^p$ tells us that events near the saturated kinematic endpoint of $\mu_{p,2}$ can be used for determination of new particle properties. 
These events have an invisible momenta solution of the corresponding $\mu_{p,2}$ approximately identical to the true invisible momenta.
The reason is that the strict convexity makes the solution unique, and thus, the true momenta are the solution when the event have the $\mu_{p,2}$ value of saturated kinematic endpoint. Like $M_{T2}$-assisted on-shell reconstruction \cite{Cho:2008tj, Choi:2009hn, Park:2011uz}, events near the saturated kinematic endpoint can be used for measuring the new particle properties.

The existence of some $\mu_{p,2}$ having saturated kinematic endpoint without the extreme kinematic conditions makes $\mu_{p,2}$ useful for mass measurement. Once a relevant $\sqrt{\hat{s}}$ is supplied, $\mu_{p,2}$ having saturated kinematic endpoint have $\sqrt{\hat{s}}$ and the upstream momentum independent endpoint $\mu_{p,2}^{\max}$ of $\mu_{p,2}$ distribution, but other endpoints of $\mu_{p,2}$ distributions deviates toward their kinematic endpoints. 
Hence, we expect such separation of kinematic endpoints in an asymmetric mass spectrum can be observed independently to the $\sqrt{\hat{s}}$ and the upstream momentum.
However, we do not know at a glance whether the kinematic endpoint of $\mu_{p,2}$ is saturated when the true mass spectrum is asymmetric, and we cannot just pick a certain $\mu_{p,2}$ to use it for mass measurement. 
Instead, for an asymmetric mass spectrum, combined analysis of $\mu_{p,2}$ distributions than analysis of $M_{T2}$ distribution only is expected to be more efficient for mass measurement.

\begin{figure}[t]
\centering
\begin{tabular}{cc}
\includegraphics[width=5.5cm]{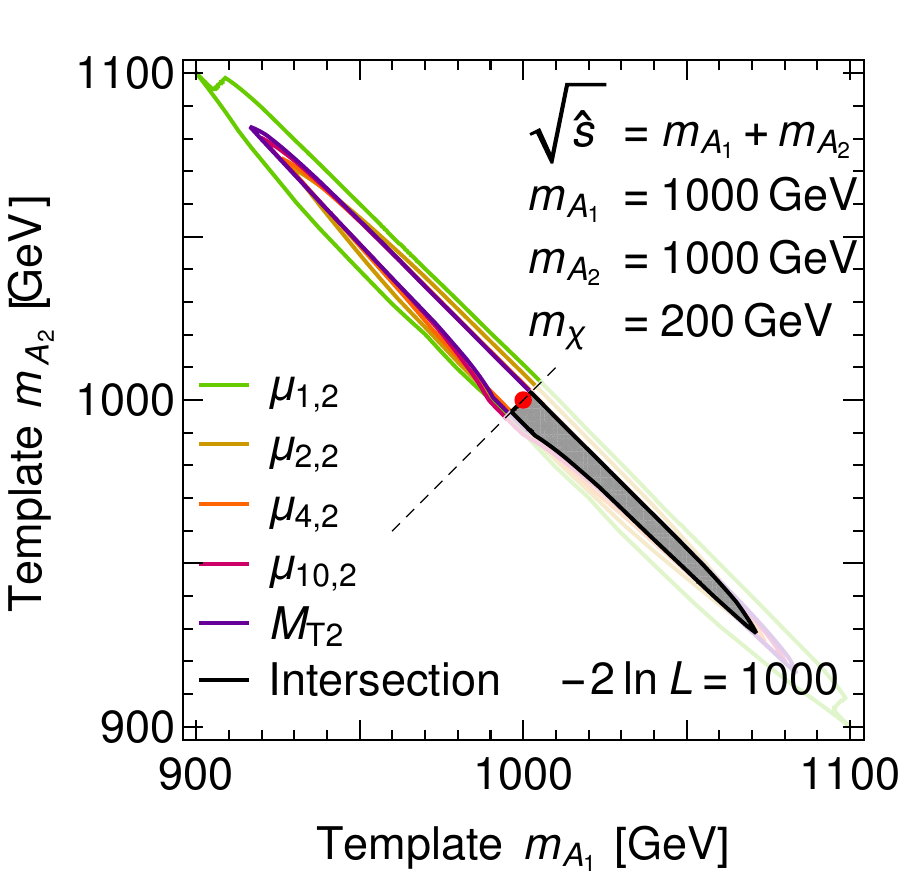} &
\includegraphics[width=5.5cm]{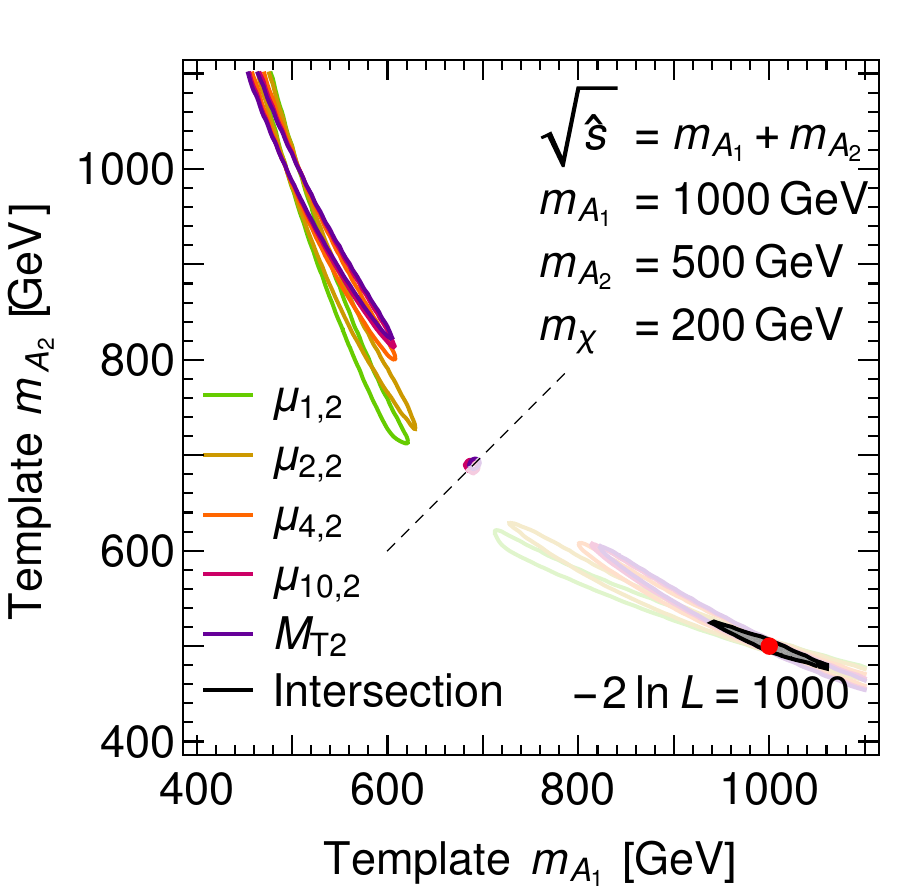} \\
\includegraphics[width=5.5cm]{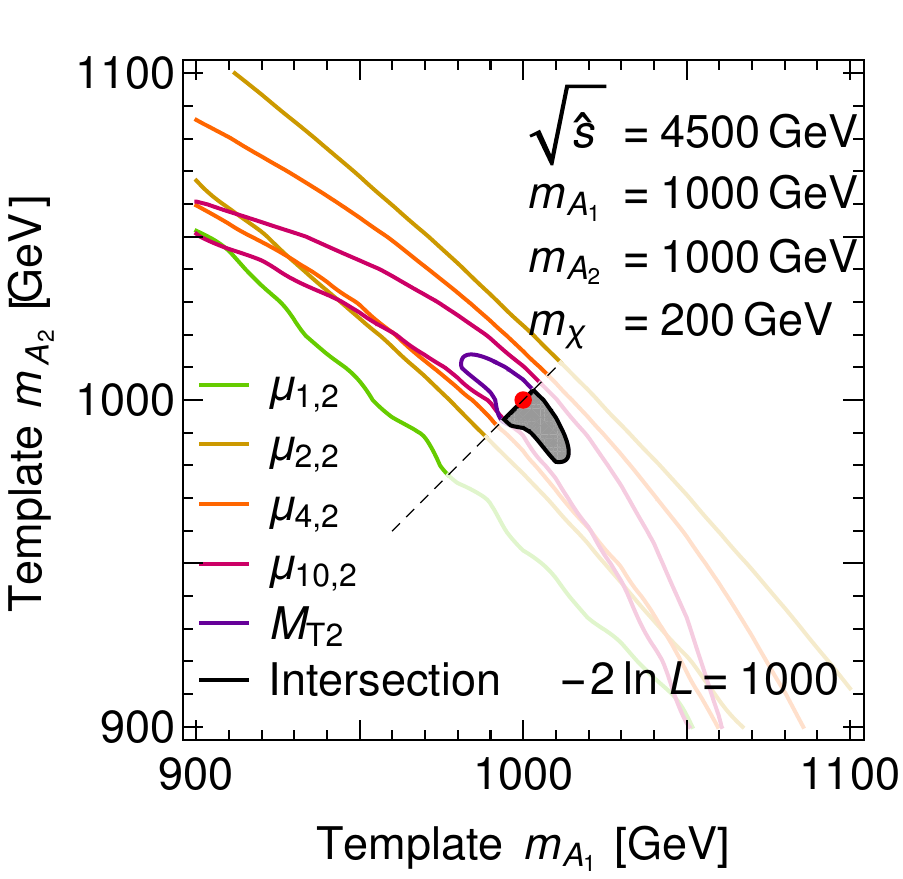} &
\includegraphics[width=5.5cm]{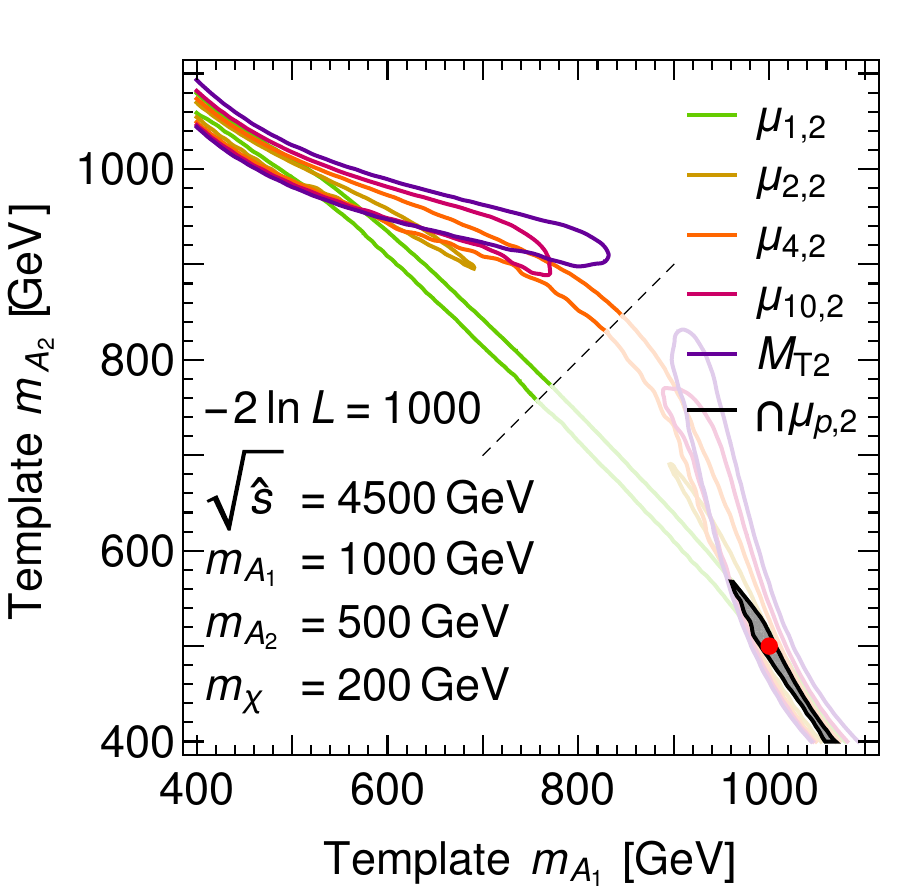} 
\end{tabular}
\caption{\label{fig:template_fit}Contours of Poisson likelihood between templates and a reference sample in a two-body decay. We generated 1,000,000 events for each template and we draw contours of $- 2 \ln L = 1000$. We compared events having at least 90\% of the $\mu_{p,2}$ value of $\mu_{p,2}^{\max}$ on a template. In the case of an asymmetric mass spectrum, the combination of $\mu_{p,2}$ shows better sensitivity in proving their masses than using $M_{T2}$ only.}
\end{figure}

Whether all of the $\mu_{p,2}$ have saturated kinematic endpoint or not, the kinematic endpoints of $\mu_{p,2}$'s are expected to have enhanced resolving power for mass spectroscopy. 
To estimate the resolving power of mass spectrum in two-body decay, we draw negative log likelihood contours, figure \ref{fig:template_fit}, using near maximal regions of $\mu_{p,2}$ distributions for a reference mass spectrum and other template mass spectrum.
To ignore a sampling fluctuation, we generated a large number of events, 1,000,000 events, for each template and we draw contours of large negative log likelihood value, $- 2 \ln L = 1000$. 
We compared events having at least 90\% of $\mu_{p,2}$ value of $\mu_{p,2}^{\max}$ on a template.
In the case of symmetric mass spectrum, $M_{T2}$ only analysis works well because there are many events near the kinematic endpoints. 
Templates with asymmetric mass spectrum typically have less saturated kinematic endpoint of $M_{T2}$ and hence likelihood can distinguish them by difference of number of events near the kinematic endpoint. 
In the case of asymmetric mass spectrum, the similar argument holds for $\mu_{p,2}$. 
However, we do not know which $\mu_{p,2}$ has a saturated kinematic endpoint, and hence, the intersection of likelihood contours of several $\mu_{p,2}$'s constrains mass spectrum better than a single likelihood contour of $\mu_{p,2}$. \\

%
\section{Comparison to Other Extensions of $M_{T2}$}
%
While in the introduction, we mentioned two non-trivial extensions of $M_{T2}$ for asymmetric parent masses, such as $M_{T2}$ with a ratio of parents masses \cite{Barr:2009jv} and subsystem $M_{T2}$ \cite{Nojiri:2008vq}. These extensions inherit property of $M_{T2}$. Hence, those mass variables use their kinematic endpoints for identifying mass spectrum, and they inherit the weak points of $M_{T2}$. We will review those methods and their key points briefly.

\subsection*{Subsystem $M_{T2}$ for an Asymmetric Mass Spectrum}

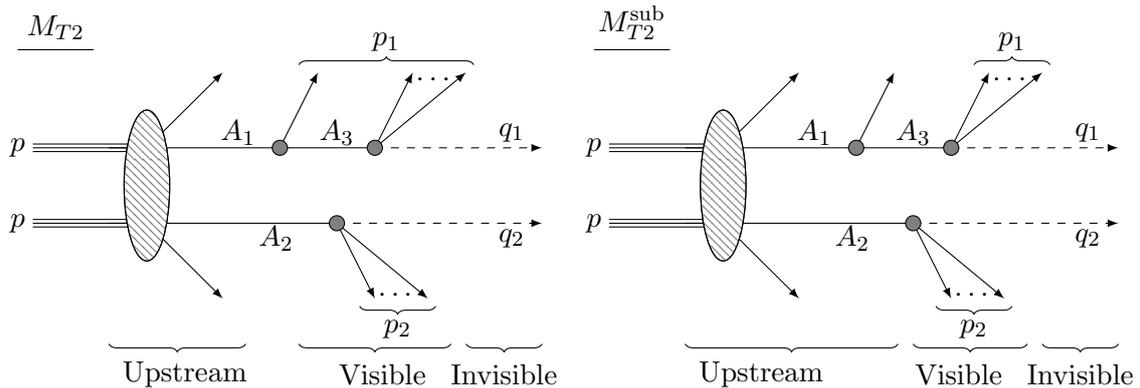
\begin{figure}[t]
\begin{center}
\begin{tikzpicture}
\draw (-2.2,1.8) -- (-1.2,1.8);
\node[draw=none,anchor=south] at (-1.7,1.8) {$M_{T2}$};
\draw (-0.5,0.5) -- (-2,0.5);
\draw (-0.5,0.55) -- (-2,0.55);
\draw (-0.5,0.45) -- (-2,0.45);
\node[draw=none] at (-2.2,0.5) {$p$};
\draw (-0.5,-0.5) -- (-2,-0.5);
\draw (-0.5,-0.55) -- (-2,-0.55);
\draw (-0.5,-0.45) -- (-2,-0.45);
\node[draw=none] at (-2.2,-0.5) {$p$};

\draw[->,>=latex] (-0.5,0.5) -- (0.5,1.5);
\draw[->,>=latex] (-0.5,-0.5) -- (0.5,-1.5);

\draw (-1,0.5) -- (2.5,0.5);
\draw (-1,-0.5) -- (2,-0.5);
\node[draw=none] at (0.7,0.7) {$A_1$};
\node[draw=none] at (1.2,-0.7) {$A_2$};
\node[draw=none] at (2.0,0.7) {$A_3$};

\draw[->,>=latex,dashed] (2.5,0.5) -- (4.7,0.5);
\draw[->,>=latex,dashed] (2,-0.5) -- (4.7,-0.5);
\node[draw=none] at (4.3,0.7) {$q_1$};
\node[draw=none] at (4.3,-0.7) {$q_2$};

\draw[->,>=latex] (1.25,0.5) -- (1.75,1.5);

\draw[->,>=latex] (2.5,0.5) -- (3.0,1.5);
\draw[->,>=latex] (2.5,0.5) -- (3.7,1.5);

\draw[->,>=latex] (2,-0.5) -- (2.5,-1.5);
\draw[->,>=latex] (2,-0.5) -- (3.2,-1.5);

\draw[decorate,decoration={brace,amplitude=3pt}] 
    (1.5, 1.6) -- 
    (3.8, 1.6) ; 
\draw[decorate,decoration={brace,amplitude=3pt,mirror}] 
    (2.3, -1.6) -- 
    (3.3, -1.6) ; 
\node[draw=none] at (2.65,1.9) {$p_1$};
\node[draw=none] at (2.8,-1.9) {$p_2$};
\node[draw=none] at (3.3,1.4) {$\cdots$};
\node[draw=none] at (2.8,-1.45) {$\cdots$};

\filldraw [color=black, fill=gray] (1.25,0.5) circle (0.1);
\filldraw [color=black, fill=gray] (2.5,0.5) circle (0.1);
\filldraw [color=black, fill=gray] (2,-0.5) circle (0.1);

\node[draw=none] at (0.0,-2.5) {Upstream};
\draw[decorate,decoration={brace,amplitude=3pt,mirror}] 
    (-1, -2.1) -- 
    (0.8,-2.1) ; 

\filldraw [color=black, fill=white] (-0.5,0) ellipse (0.3 and 1);
\draw[pattern=north west lines, pattern color=gray] (-0.5,0) ellipse (0.3 and 1);

\node[draw=none] at (2.6,-2.5) {Visible};
\draw[decorate,decoration={brace,amplitude=3pt,mirror}] 
    (1.5, -2.1) -- 
    (3.5,-2.1) ; 
    
\node[draw=none] at (4.2,-2.5) {Invisible};
\draw[decorate,decoration={brace,amplitude=3pt,mirror}] 
    (3.7, -2.1) -- 
    (4.7,-2.1) ;  
\end{tikzpicture}
\begin{tikzpicture}
\draw (-2.2,1.8) -- (-1.2,1.8);
\node[draw=none,anchor=south] at (-1.7,1.8) {$M_{T2}^{\mathrm{sub}}$};
\draw (-0.5,0.5) -- (-2,0.5);
\draw (-0.5,0.55) -- (-2,0.55);
\draw (-0.5,0.45) -- (-2,0.45);
\node[draw=none] at (-2.2,0.5) {$p$};
\draw (-0.5,-0.5) -- (-2,-0.5);
\draw (-0.5,-0.55) -- (-2,-0.55);
\draw (-0.5,-0.45) -- (-2,-0.45);
\node[draw=none] at (-2.2,-0.5) {$p$};

\draw[->,>=latex] (-0.5,0.5) -- (0.5,1.5);
\draw[->,>=latex] (-0.5,-0.5) -- (0.5,-1.5);

\draw (-1,0.5) -- (2.5,0.5);
\draw (-1,-0.5) -- (2,-0.5);
\node[draw=none] at (0.7,0.7) {$A_1$};
\node[draw=none] at (1.2,-0.7) {$A_2$};
\node[draw=none] at (2.0,0.7) {$A_3$};

\draw[->,>=latex,dashed] (2.5,0.5) -- (4.7,0.5);
\draw[->,>=latex,dashed] (2,-0.5) -- (4.7,-0.5);
\node[draw=none] at (4.3,0.7) {$q_1$};
\node[draw=none] at (4.3,-0.7) {$q_2$};

\draw[->,>=latex] (1.25,0.5) -- (1.75,1.5);

\draw[->,>=latex] (2.5,0.5) -- (3.0,1.5);
\draw[->,>=latex] (2.5,0.5) -- (3.7,1.5);

\draw[->,>=latex] (2,-0.5) -- (2.5,-1.5);
\draw[->,>=latex] (2,-0.5) -- (3.2,-1.5);

\draw[decorate,decoration={brace,amplitude=3pt}] 
    (2.8, 1.6) -- 
    (3.8, 1.6) ; 
\draw[decorate,decoration={brace,amplitude=3pt,mirror}] 
    (2.3, -1.6) -- 
    (3.3, -1.6) ; 
\node[draw=none] at (3.3,1.9) {$p_1$};
\node[draw=none] at (2.8,-1.9) {$p_2$};
\node[draw=none] at (3.3,1.4) {$\cdots$};
\node[draw=none] at (2.8,-1.45) {$\cdots$};

\filldraw [color=black, fill=gray] (1.25,0.5) circle (0.1);
\filldraw [color=black, fill=gray] (2.5,0.5) circle (0.1);
\filldraw [color=black, fill=gray] (2,-0.5) circle (0.1);

\node[draw=none] at (0.0,-2.5) {Upstream};
\draw[decorate,decoration={brace,amplitude=3pt,mirror}] 
    (-1, -2.1) -- 
    (1.8,-2.1) ; 

\filldraw [color=black, fill=white] (-0.5,0) ellipse (0.3 and 1);
\draw[pattern=north west lines, pattern color=gray] (-0.5,0) ellipse (0.3 and 1);

\node[draw=none] at (2.6,-2.5) {Visible};
\draw[decorate,decoration={brace,amplitude=3pt,mirror}] 
    (2.0, -2.1) -- 
    (3.5,-2.1) ; 
    
\node[draw=none] at (4.2,-2.5) {Invisible};
\draw[decorate,decoration={brace,amplitude=3pt,mirror}] 
    (3.7, -2.1) -- 
    (4.7,-2.1) ;  
\end{tikzpicture}
\end{center}
\caption{\label{fig:process_topology_subsystem} A pair production of two resonances $A_1$ and $A_2$ having masses $m_{A_1}$ and $m_{A_2}$ decaying semi-invisibly in a collider. $A_1$ is decaying to a visible particle and $A_3$ first, and after then $A_3$ decays semi-invisibly. In $M_{T2}$, we will consider all visible particle produced from decay of $A_1$ and $A_2$. While in $M_{T2}^{\mathrm{sub}}$, we discard the visible particle produced from decay of $A_1$. }
\end{figure}

If there is a sub-resonance inside of a decay chain of heavier resonance, we can utilize the sub-resonance for identifying mass of the lighter resonance. 
Let us consider a pair production of $A_1$ and $A_2$ while $A_1$ has a sub-resonance $A_3$ in its decay chain.
In this case, we can think two kinds of $M_{T2}$. 
One is $M_{T2}$ constructed from $A_1$ and $A_2$, and the other is $M_{T2}$ constructed from the subsystem $A_3$ and $A_2$. 
The second one is called \emph{subsystem $M_{T2}$} \cite{Nojiri:2008vq}, and we denote it as $M_{T2}^{\mathrm{sub}}$. 
We illustrate systems where each $M_{T2}$ considers in figure  \ref{fig:process_topology_subsystem}.
Those two $M_{T2}$'s have upper bounds,
\begin{eqnarray}
M_{T2}(m_\chi) 
& \leq &
\max( m_{A_1} , m_{A_2} ) 
\\
M_{T2}^{\mathrm{sub}}(m_\chi) 
& \leq &
\max( m_{A_2} , m_{A_3} ) 
\end{eqnarray}
If these upper bounds are saturated, then we can identify at least two of resonance masses in this process.

An example for this event topology is a gluino-squark coproduction with mass hierarchy $m_{\tilde{q}} > m_{\tilde{g}}$ \cite{Nojiri:2008vq}. In this event topology, $A_1$ is a squark $\tilde{q}$, and $A_2$ and $A_3$ are gluinos $\tilde{g}$. We assume that the squark decays to a quark $q$ and a gluino $\tilde{g}$, and two gluinos decay to a quark-antiquark pair and a gluon. Then the upper bound of $M_{T2}$ and $M_{T2}^{\mathrm{sub}}$ reveals masses of squark $m_{\tilde{q}}$ and gluino $m_{\tilde{g}}$.
\begin{eqnarray}
M_{T2}(m_\chi) 
& \leq &
\max( m_{\tilde{q}} , m_{\tilde{g}} )  = m_{\tilde{q}}
\\
M_{T2}^{\mathrm{sub}}(m_\chi) 
& \leq &
\max( m_{\tilde{g}} , m_{\tilde{g}} )  = m_{\tilde{g}}
\end{eqnarray}
These upper bounds are saturated, and we can use the endpoints for mass measurement \cite{Nojiri:2008vq}.

One difficulty of this method is that we need to identify visible particles produced in the decay of $A_1$ to clarify the subsystem $A_2$ and $A_3$. While in the gluino-squark coproduction, if the mass gap $m_{\tilde{q}} - m_{\tilde{g}}$ is large enough, the decay of $\tilde{q}$ produce a high $p_T$ quark $q$. Hence, we can discard a high $p_T$ jet, and we can evaluate $M_{T2}^{\mathrm{sub}}$ without a combinatorial ambiguity.

Moreover, there would be an ambiguity about the relation between $m_{A_2}$ and $m_{A_3}$ without a model assumption, because the kinematic endpoint only says $\max (m_{A_2}, m_{A_3})$.
The gluinos are decaying to three-body, and hence, $M_{T2}^{\mathrm{sub}}$ saturates its upper bound also when $m_{A_2} \neq m_{A_3}$ as figure \ref{fig:sample_distribution_threebody_threshold}. 
To resolve this problem, we can look for kinematic endpoints of invariant masses of two jets produced by $A_2$ or $A_3$, or we can look for the kinematic endpoint of $\mu_{p,2}$  to solve this ambiguity, especially when $A_2$ and $A_3$ are decaying to two-body.

\subsection*{$M_{T2}$ with mass ratio}
Instead of finding a subsystem, we can alter the definition of $M_{T2}$ to make it more suitable for measuring non-identical masses of parents particle. One reason why $M_{T2}$ works well with a symmetric mass spectrum is that its implicit constraint, $M_{A_1} = M_{A_2}$, is compatible with the true mass spectrum. We can modify this constraint by weighting $M_{A_i}$ by a hypothetical mass ratio $\tilde{m}_{A_2} / \tilde{m}_{A_1}$ of particles $A_1$ and $A_2$. This procedure suggests a generalized $M_{T2}$ \cite{Barr:2009jv}, 
\begin{equation}
\label{eqn:definition_M2_weighted}
M_{T2}(\tilde{m};\tilde{m}_{A_1} / \tilde{m}_{A_2})  = \min_{\substack{
	\mathbf{q}_{1}, \mathbf{q}_{2} \\ 
	q_1^a + q_2^a = \slashed{p} {}_T^a
}}   \max \left( \sqrt{\frac{\tilde{m}_{A_2}}{\tilde{m}_{A_1}}}
M_{A_1} \left( p_1^\mu, q_1^\mu(\tilde{m}) \right) 
, \sqrt{\frac{\tilde{m}_{A_1}}{\tilde{m}_{A_2}}} 
M_{A_2} \left( p_2^\mu, q_2^\mu(\tilde{m}) \right) 
\right) \;.
\end{equation}
We can replace each $M_{A_i}$ to the transverse mass $M_{T,A_i}$ after the minimization on longitudinal components. Since a balanced solution of this $M_{T2}$ is located where arguments of the maximum function are identical, we have an implicit constraint,
\begin{equation}\label{eqn:MT2_Weighted_implicit_constarint}
\frac{M_{A_1}}{\tilde{m}_{A_1}} = \frac{M_{A_2}}{\tilde{m}_{A_2}} \,.
\end{equation}
Therefore, if the true mass spectrum is compatible to the hypothetical mass ratio $\tilde{m}_{A_1} / \tilde{m}_{A_2}$, then we expect that this generalized $M_{T2}$ works better than the usual $M_{T2}$.

\begin{figure}[t]
\centering
\begin{tabular}{ccc}
\includegraphics[width=4.6cm]{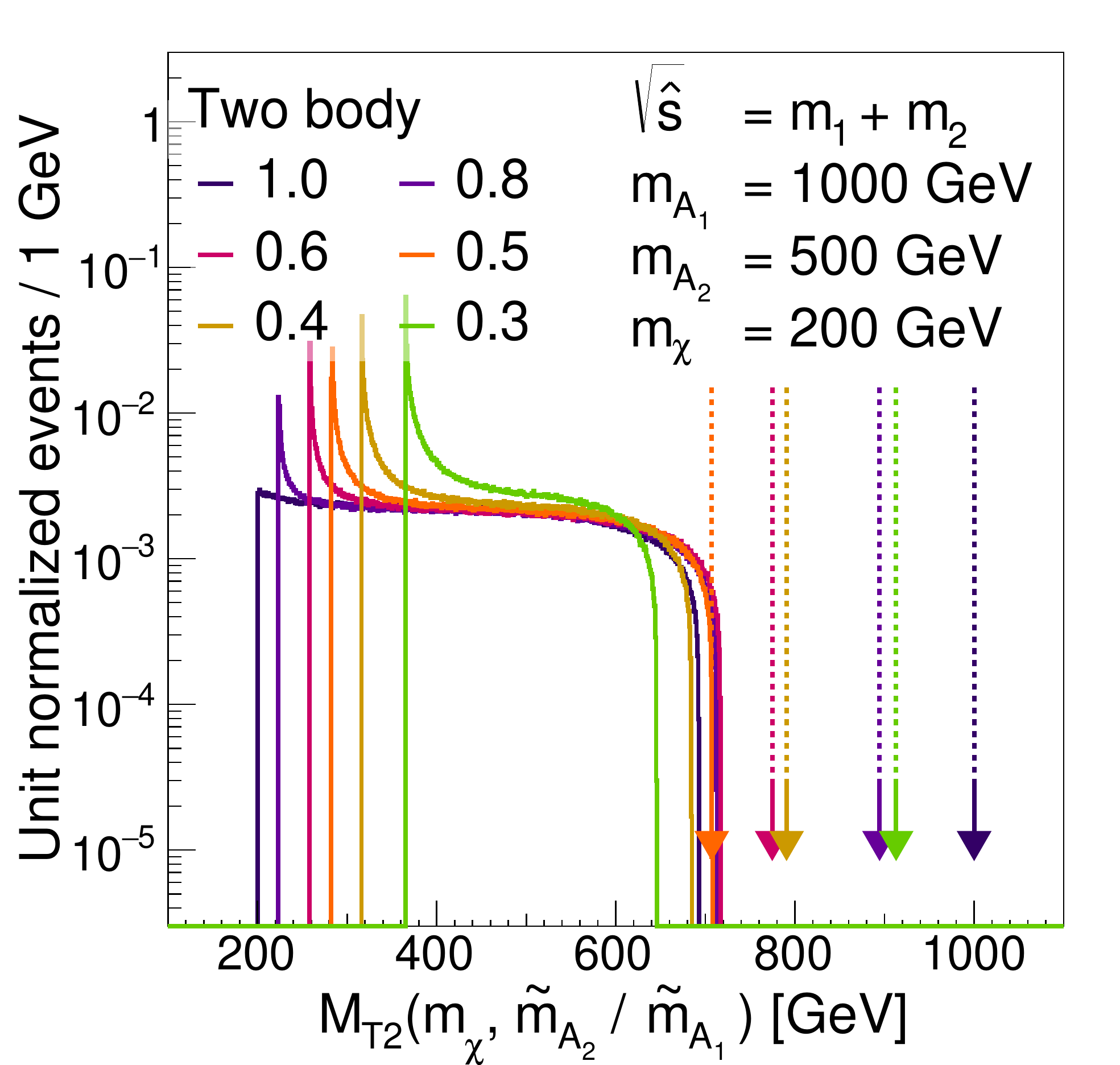} &
\includegraphics[width=4.6cm]{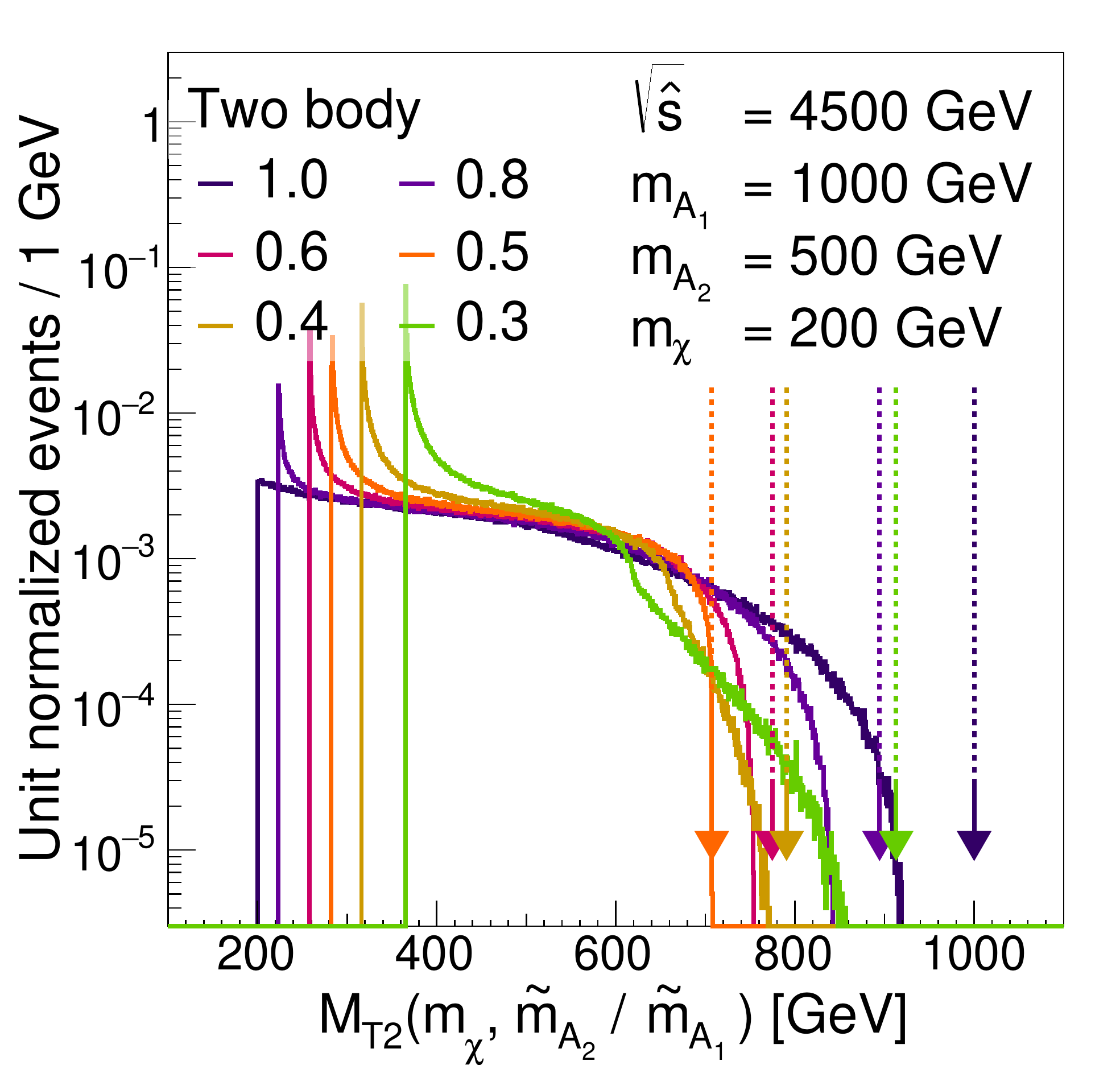} &
\includegraphics[width=4.6cm]{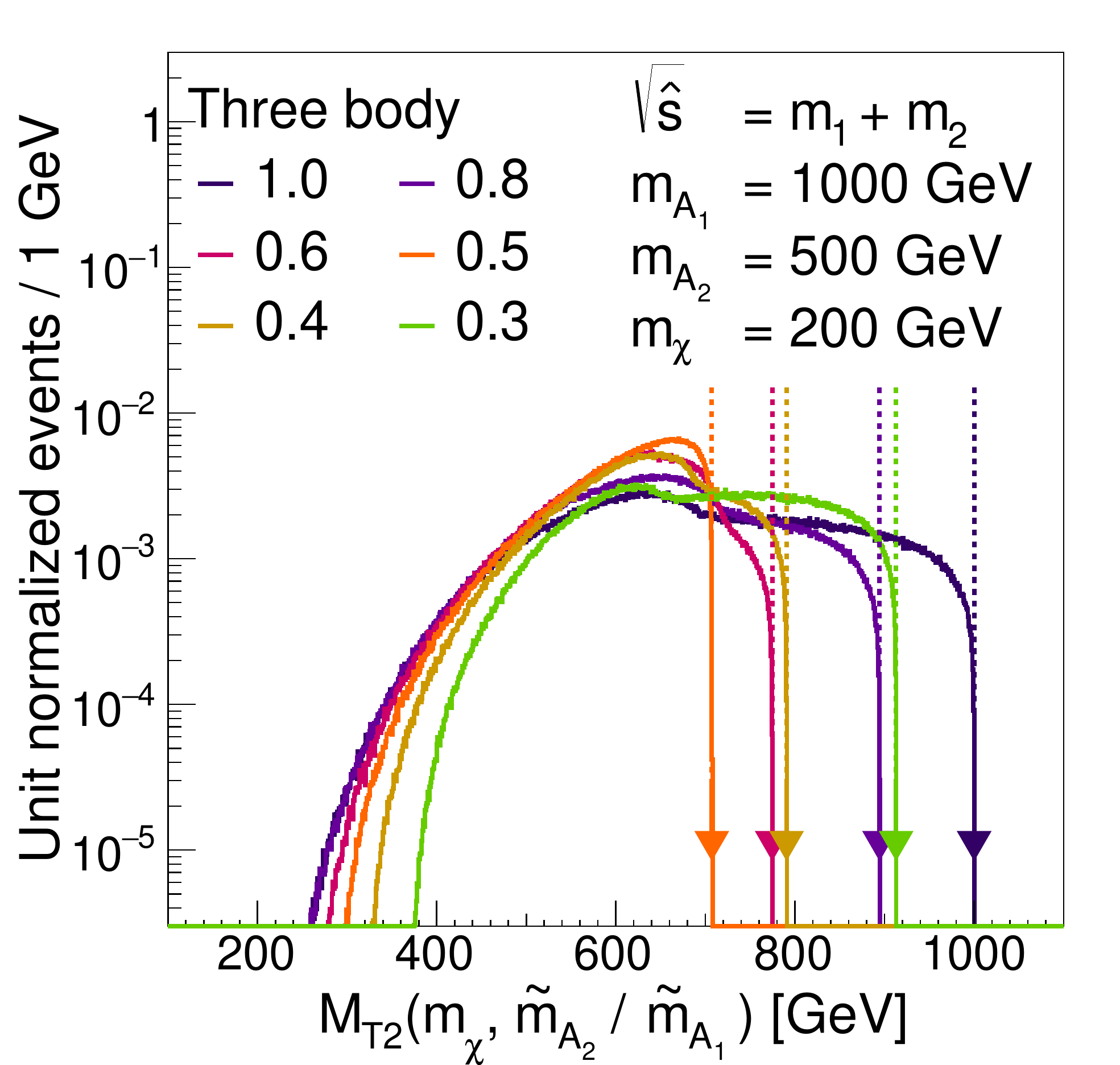} 
\end{tabular}
\caption{\label{fig:sample_distribution_MT2_weighted}Mass-ratio weighted $M_{T2}$ distribution in two-body decay and three-body decay for a threshold production.  Arrows show the locations of the upper bounds for each distribution. Legends on each line indicate mass ratios $\tilde{m}_{A_2} / \tilde{m}_{A_1}$. In the case of two-body decay, the upper bound is saturated only when the mass ratio is the true mass ratio.}
\end{figure}

By its method of construction, this generalized $M_{T2}$ has upper and lower bounds, 
\begin{eqnarray}
M_{T2}(m_\chi;m_{A_2} / m_{A_1}) 
& \leq &
\sqrt{ m_{A_1} m_{A_2}}  \;, 
\\
M_{T2}(m_\chi;\tilde{m}_{A_2} / \tilde{m}_{A_1}) 
& \leq &
\max \left( 
\sqrt{\frac{\tilde{m}_{A_2}}{\tilde{m}_{A_1}}}
m_{A_1}, 
\sqrt{\frac{\tilde{m}_{A_1}}{\tilde{m}_{A_2}}} 
m_{A_2}
\right) \;,
\\
M_{T2}(\tilde{m};\tilde{m}_{A_2} / \tilde{m}_{A_1}) 
& \geq &
\max \left( 
\sqrt{\frac{\tilde{m}_{A_2}}{\tilde{m}_{A_1}}}
(m_{p_1} + \tilde{m}), 
\sqrt{\frac{\tilde{m}_{A_1}}{\tilde{m}_{A_2}}} 
(m_{p_2} + \tilde{m})
\right) \;,
\end{eqnarray}
Because of similar reasoning in section 3, we expect that these upper bounds are saturated when the implicit constraint in eq. \ref{eqn:MT2_Weighted_implicit_constarint} is compatible to the true mass spectrum. In the three-body decay case, these upper bound can saturate upper bounds regardless of the hypothetical mass ratio $\tilde{m}_{A_2} / \tilde{m}_{A_1}$, because the lower bound pushes forward $M_{T2}$ value to the upper bound. We illustrate this endpoint behavior in figure \ref{fig:sample_distribution_MT2_weighted}. 

The strong point of this method is that extreme kinematic conditions are not required if we used a correct mass ratio. As we illustrated in figure \ref{fig:sample_distribution_twobody_threshold}, if we are going to measure mass spectrum by the kinematic endpoints, not by comparing to templates, then the usual $M_{T2}$ requires extreme kinematic conditions, and $\mu_{p,2}$ requires some moderate kinematic conditions. If one can find out the true mass ratio $m_{A_2} / m_{A_1}$, then this generalized $M_{T2}$ is a good method for mass measurement.

However, the mass ratio $m_{A_2} / m_{A_1}$ is unknown at the beginning, and hence, we should interpret the kinematic endpoint cautiously. 
As we illustrated in figure \ref{fig:sample_distribution_MT2_weighted}, using wrong $\tilde{m}_{A_2} / \tilde{m}_{A_1}$ can give us a wrong conclusion because the kinematic endpoint does not reach the upper bound especially in two-body decay case. To use this generalized $M_{T2}$ for mass measurement, choosing a correct $\tilde{m}_{A_2} / \tilde{m}_{A_1}$ is very important. 
We may leave this problem by comparing data to templates as mentioned in \cite{Barr:2009jv} at a cost of model-dependent interpretation of the data. Without a template or a model assumption, we cannot interpret the kinematic endpoint of the generalized $M_{T2}$ directly, because we don't know that used $\tilde{m}_{A_2} / \tilde{m}_{A_1}$ is correct or not. Meanwhile, the kinematic endpoints of $\mu_{p,2}$'s at least tell us that the true mass spectrum is symmetric or asymmetric.

%
\section{Conclusions}
%
In this work, we have developed new kinematic variables $\mu_{p,2}$, which is a smooth generalization of $M_{T2}$ and $M_{2}$ variable, for measuring masses of pair-produced particles decaying semi-invisibly.
$\mu_{p,2}$'s have distinctive kinematic endpoints for symmetric and asymmetric mass spectrum, and it can be used for surveying the true mass spectrum. 
If the true mass spectrum is asymmetric, $\mu_{p,2}$'s have separated kinematic endpoints, whereas if the true mass spectrum is symmetric, $\mu_{p,2}$'s share common kinematic endpoints.
While $M_{T2}$ analysis has several foibles, which are only larger resonance mass dependence of the upper bound and weak saturation of kinematic endpoint when the true mass spectrum is asymmetric, combined analysis of $\mu_{p,2}$ constrain the true mass spectrum better.
We expect $\mu_{p,2}$ is particularly useful for mass determination of new physics particles when underlying physics is undetermined, such as slepton pair production and chargino-neutralino co-production.
Although saturation of the kinematic endpoints of $\mu_{p,2}$'s depends on event topology, likelihood analysis shows that mass measurement by $\mu_{p,2}$ can be promising.
We further expect there should be a kinematic method for quantifying asymmetric true mass spectrum in the case that only some of $\mu_{p,2}$'s have saturated kinematic endpoints, and further study is ongoing \cite{shlim:next}.
While in this paper, an effect of wrong invisible particle mass is not discussed and it is not ignorable in typical new physics scenarios. Further study on the relationship between trial invisible particle mass and the kinematic endpoint of $\mu_{p,2}$ is needed.

%
\acknowledgments
%
I would like to thank K. Choi for supporting and general advice.
This work was supported by IBS under the project code, IBS-R018-D1.

\bibliographystyle{JHEP}
\bibliography{PowerMean}

\providecommand{\href}[2]{#2}\begingroup\raggedright\begin{thebibliography}{10}

\bibitem{Feng:2010gw}
J.~L. Feng, \emph{{Dark Matter Candidates from Particle Physics and Methods of
  Detection}},
  \href{http://dx.doi.org/10.1146/annurev-astro-082708-101659}{\emph{Ann. Rev.
  Astron. Astrophys.} {\bf 48} (2010) 495--545},
  [\href{http://arxiv.org/abs/1003.0904}{{\tt 1003.0904}}].

\bibitem{Nilles:1983ge}
H.~P. Nilles, \emph{{Supersymmetry, Supergravity and Particle Physics}},
  \href{http://dx.doi.org/10.1016/0370-1573(84)90008-5}{\emph{Phys. Rept.} {\bf
  110} (1984) 1--162}.

\bibitem{Haber:1984rc}
H.~E. Haber and G.~L. Kane, \emph{{The Search for Supersymmetry: Probing
  Physics Beyond the Standard Model}},
  \href{http://dx.doi.org/10.1016/0370-1573(85)90051-1}{\emph{Phys. Rept.} {\bf
  117} (1985) 75--263}.

\bibitem{Appelquist:2000nn}
T.~Appelquist, H.-C. Cheng and B.~A. Dobrescu, \emph{{Bounds on universal extra
  dimensions}}, \href{http://dx.doi.org/10.1103/PhysRevD.64.035002}{\emph{Phys.
  Rev.} {\bf D64} (2001) 035002},
  [\href{http://arxiv.org/abs/hep-ph/0012100}{{\tt hep-ph/0012100}}].

\bibitem{Kim:2009si}
I.-W. Kim, \emph{{Algebraic Singularity Method for Mass Measurement with
  Missing Energy}},
  \href{http://dx.doi.org/10.1103/PhysRevLett.104.081601}{\emph{Phys. Rev.
  Lett.} {\bf 104} (2010) 081601}, [\href{http://arxiv.org/abs/0910.1149}{{\tt
  0910.1149}}].

\bibitem{Hinchliffe:1996iu}
I.~Hinchliffe, F.~E. Paige, M.~D. Shapiro, J.~Soderqvist and W.~Yao,
  \emph{{Precision SUSY measurements at CERN LHC}},
  \href{http://dx.doi.org/10.1103/PhysRevD.55.5520}{\emph{Phys. Rev.} {\bf D55}
  (1997) 5520--5540}, [\href{http://arxiv.org/abs/hep-ph/9610544}{{\tt
  hep-ph/9610544}}].

\bibitem{Hinchliffe:1999zc}
I.~Hinchliffe and F.~E. Paige, \emph{{Measurements in SUGRA models with large
  tan beta at CERN LHC}},
  \href{http://dx.doi.org/10.1103/PhysRevD.61.095011}{\emph{Phys. Rev.} {\bf
  D61} (2000) 095011}, [\href{http://arxiv.org/abs/hep-ph/9907519}{{\tt
  hep-ph/9907519}}].

\bibitem{Tovey:2008ui}
D.~R. Tovey, \emph{{On measuring the masses of pair-produced semi-invisibly
  decaying particles at hadron colliders}},
  \href{http://dx.doi.org/10.1088/1126-6708/2008/04/034}{\emph{JHEP} {\bf 04}
  (2008) 034}, [\href{http://arxiv.org/abs/0802.2879}{{\tt 0802.2879}}].

\bibitem{Han:2009ss}
T.~Han, I.-W. Kim and J.~Song, \emph{{Kinematic Cusps: Determining the Missing
  Particle Mass at Colliders}},
  \href{http://dx.doi.org/10.1016/j.physletb.2010.09.010}{\emph{Phys. Lett.}
  {\bf B693} (2010) 575--579}, [\href{http://arxiv.org/abs/0906.5009}{{\tt
  0906.5009}}].

\bibitem{Barr:2010zj}
A.~J. Barr and C.~G. Lester, \emph{{A Review of the Mass Measurement Techniques
  proposed for the Large Hadron Collider}},
  \href{http://dx.doi.org/10.1088/0954-3899/37/12/123001}{\emph{J. Phys.} {\bf
  G37} (2010) 123001}, [\href{http://arxiv.org/abs/1004.2732}{{\tt
  1004.2732}}].

\bibitem{Cho:2012er}
W.~S. Cho, D.~Kim, K.~T. Matchev and M.~Park, \emph{{Probing Resonance Decays
  to Two Visible and Multiple Invisible Particles}},
  \href{http://dx.doi.org/10.1103/PhysRevLett.112.211801}{\emph{Phys. Rev.
  Lett.} {\bf 112} (2014) 211801}, [\href{http://arxiv.org/abs/1206.1546}{{\tt
  1206.1546}}].

\bibitem{Kim:2015bnd}
D.~Kim, K.~T. Matchev and M.~Park, \emph{{Using sorted invariant mass variables
  to evade combinatorial ambiguities in cascade decays}},
  \href{http://dx.doi.org/10.1007/JHEP02(2016)129}{\emph{JHEP} {\bf 02} (2016)
  129}, [\href{http://arxiv.org/abs/1512.02222}{{\tt 1512.02222}}].

\bibitem{Lester:1999tx}
C.~G. Lester and D.~J. Summers, \emph{{Measuring masses of semiinvisibly
  decaying particles pair produced at hadron colliders}},
  \href{http://dx.doi.org/10.1016/S0370-2693(99)00945-4}{\emph{Phys. Lett.}
  {\bf B463} (1999) 99--103}, [\href{http://arxiv.org/abs/hep-ph/9906349}{{\tt
  hep-ph/9906349}}].

\bibitem{Barr:2003rg}
A.~Barr, C.~Lester and P.~Stephens, \emph{{m(T2): The Truth behind the
  glamour}}, \href{http://dx.doi.org/10.1088/0954-3899/29/10/304}{\emph{J.
  Phys.} {\bf G29} (2003) 2343--2363},
  [\href{http://arxiv.org/abs/hep-ph/0304226}{{\tt hep-ph/0304226}}].

\bibitem{Cho:2007qv}
W.~S. Cho, K.~Choi, Y.~G. Kim and C.~B. Park, \emph{{Gluino Stransverse Mass}},
  \href{http://dx.doi.org/10.1103/PhysRevLett.100.171801}{\emph{Phys. Rev.
  Lett.} {\bf 100} (2008) 171801}, [\href{http://arxiv.org/abs/0709.0288}{{\tt
  0709.0288}}].

\bibitem{Cheng:2008hk}
H.-C. Cheng and Z.~Han, \emph{{Minimal Kinematic Constraints and m(T2)}},
  \href{http://dx.doi.org/10.1088/1126-6708/2008/12/063}{\emph{JHEP} {\bf 12}
  (2008) 063}, [\href{http://arxiv.org/abs/0810.5178}{{\tt 0810.5178}}].

\bibitem{Barr:2011xt}
A.~J. Barr, T.~J. Khoo, P.~Konar, K.~Kong, C.~G. Lester, K.~T. Matchev et~al.,
  \emph{{Guide to transverse projections and mass-constraining variables}},
  \href{http://dx.doi.org/10.1103/PhysRevD.84.095031}{\emph{Phys. Rev.} {\bf
  D84} (2011) 095031}, [\href{http://arxiv.org/abs/1105.2977}{{\tt
  1105.2977}}].

\bibitem{Mahbubani:2012kx}
R.~Mahbubani, K.~T. Matchev and M.~Park, \emph{{Re-interpreting the Oxbridge
  stransverse mass variable MT2 in general cases}},
  \href{http://dx.doi.org/10.1007/JHEP03(2013)134}{\emph{JHEP} {\bf 03} (2013)
  134}, [\href{http://arxiv.org/abs/1212.1720}{{\tt 1212.1720}}].

\bibitem{Cho:2014naa}
W.~S. Cho, J.~S. Gainer, D.~Kim, K.~T. Matchev, F.~Moortgat, L.~Pape et~al.,
  \emph{{On-shell constrained $M_2$ variables with applications to mass
  measurements and topology disambiguation}},
  \href{http://dx.doi.org/10.1007/JHEP08(2014)070}{\emph{JHEP} {\bf 08} (2014)
  070}, [\href{http://arxiv.org/abs/1401.1449}{{\tt 1401.1449}}].

\bibitem{Cho:2015laa}
W.~S. Cho, J.~S. Gainer, D.~Kim, S.~H. Lim, K.~T. Matchev, F.~Moortgat et~al.,
  \emph{{OPTIMASS: A Package for the Minimization of Kinematic Mass Functions
  with Constraints}},
  \href{http://dx.doi.org/10.1007/JHEP01(2016)026}{\emph{JHEP} {\bf 01} (2016)
  026}, [\href{http://arxiv.org/abs/1508.00589}{{\tt 1508.00589}}].

\bibitem{Konar:2009wn}
P.~Konar, K.~Kong, K.~T. Matchev and M.~Park, \emph{{Superpartner Mass
  Measurement Technique using 1D Orthogonal Decompositions of the Cambridge
  Transverse Mass Variable $M_{T2}$}},
  \href{http://dx.doi.org/10.1103/PhysRevLett.105.051802}{\emph{Phys. Rev.
  Lett.} {\bf 105} (2010) 051802}, [\href{http://arxiv.org/abs/0910.3679}{{\tt
  0910.3679}}].

\bibitem{Cho:2009ve}
W.~S. Cho, J.~E. Kim and J.-H. Kim, \emph{{Amplification of endpoint structure
  for new particle mass measurement at the LHC}},
  \href{http://dx.doi.org/10.1103/PhysRevD.81.095010}{\emph{Phys. Rev.} {\bf
  D81} (2010) 095010}, [\href{http://arxiv.org/abs/0912.2354}{{\tt
  0912.2354}}].

\bibitem{Konar:2015hea}
P.~Konar and A.~K. Swain, \emph{{Mass reconstruction with $M_2$ under
  constraint in semi-invisible production at a hadron collider}},
  \href{http://dx.doi.org/10.1103/PhysRevD.93.015021}{\emph{Phys. Rev.} {\bf
  D93} (2016) 015021}, [\href{http://arxiv.org/abs/1509.00298}{{\tt
  1509.00298}}].

\bibitem{Gripaios:2007is}
B.~Gripaios, \emph{{Transverse observables and mass determination at hadron
  colliders}},
  \href{http://dx.doi.org/10.1088/1126-6708/2008/02/053}{\emph{JHEP} {\bf 02}
  (2008) 053}, [\href{http://arxiv.org/abs/0709.2740}{{\tt 0709.2740}}].

\bibitem{Barr:2007hy}
A.~J. Barr, B.~Gripaios and C.~G. Lester, \emph{{Weighing Wimps with Kinks at
  Colliders: Invisible Particle Mass Measurements from Endpoints}},
  \href{http://dx.doi.org/10.1088/1126-6708/2008/02/014}{\emph{JHEP} {\bf 02}
  (2008) 014}, [\href{http://arxiv.org/abs/0711.4008}{{\tt 0711.4008}}].

\bibitem{Cho:2007dh}
W.~S. Cho, K.~Choi, Y.~G. Kim and C.~B. Park, \emph{{Measuring superparticle
  masses at hadron collider using the transverse mass kink}},
  \href{http://dx.doi.org/10.1088/1126-6708/2008/02/035}{\emph{JHEP} {\bf 02}
  (2008) 035}, [\href{http://arxiv.org/abs/0711.4526}{{\tt 0711.4526}}].

\bibitem{Konar:2009qr}
P.~Konar, K.~Kong, K.~T. Matchev and M.~Park, \emph{{Dark Matter Particle
  Spectroscopy at the LHC: Generalizing M(T2) to Asymmetric Event Topologies}},
  \href{http://dx.doi.org/10.1007/JHEP04(2010)086}{\emph{JHEP} {\bf 04} (2010)
  086}, [\href{http://arxiv.org/abs/0911.4126}{{\tt 0911.4126}}].

\bibitem{Nojiri:2008hy}
M.~M. Nojiri, Y.~Shimizu, S.~Okada and K.~Kawagoe, \emph{{Inclusive transverse
  mass analysis for squark and gluino mass determination}},
  \href{http://dx.doi.org/10.1088/1126-6708/2008/06/035}{\emph{JHEP} {\bf 06}
  (2008) 035}, [\href{http://arxiv.org/abs/0802.2412}{{\tt 0802.2412}}].

\bibitem{Barr:2009wu}
A.~J. Barr and C.~Gwenlan, \emph{{The Race for supersymmetry: Using m(T2) for
  discovery}}, \href{http://dx.doi.org/10.1103/PhysRevD.80.074007}{\emph{Phys.
  Rev.} {\bf D80} (2009) 074007}, [\href{http://arxiv.org/abs/0907.2713}{{\tt
  0907.2713}}].

\bibitem{Cho:2008cu}
W.~S. Cho, K.~Choi, Y.~G. Kim and C.~B. Park, \emph{{Measuring the top quark
  mass with m(T2) at the LHC}},
  \href{http://dx.doi.org/10.1103/PhysRevD.78.034019}{\emph{Phys. Rev.} {\bf
  D78} (2008) 034019}, [\href{http://arxiv.org/abs/0804.2185}{{\tt
  0804.2185}}].

\bibitem{Aaltonen:2009rm}
{\scshape CDF} collaboration, T.~Aaltonen et~al., \emph{{Measurement of the Top
  Quark Mass in the Dilepton Channel Using mT2 at CDF}},
  \href{http://dx.doi.org/10.1103/PhysRevD.81.031102}{\emph{Phys. Rev.} {\bf
  D81} (2010) 031102}, [\href{http://arxiv.org/abs/0911.2956}{{\tt
  0911.2956}}].

\bibitem{Chatrchyan:2013boa}
{\scshape CMS} collaboration, S.~Chatrchyan et~al., \emph{{Measurement of
  masses in the $t \bar{t}$ system by kinematic endpoints in pp collisions at
  $\sqrt{s}$ = 7 TeV}},
  \href{http://dx.doi.org/10.1140/epjc/s10052-013-2494-7}{\emph{Eur. Phys. J.}
  {\bf C73} (2013) 2494}, [\href{http://arxiv.org/abs/1304.5783}{{\tt
  1304.5783}}].

\bibitem{Barr:2009jv}
A.~J. Barr, B.~Gripaios and C.~G. Lester, \emph{{Transverse masses and
  kinematic constraints: from the boundary to the crease}},
  \href{http://dx.doi.org/10.1088/1126-6708/2009/11/096}{\emph{JHEP} {\bf 11}
  (2009) 096}, [\href{http://arxiv.org/abs/0908.3779}{{\tt 0908.3779}}].

\bibitem{Nojiri:2008vq}
M.~M. Nojiri, K.~Sakurai, Y.~Shimizu and M.~Takeuchi, \emph{{Handling jets +
  missing E(T) channel using inclusive m(T2)}},
  \href{http://dx.doi.org/10.1088/1126-6708/2008/10/100}{\emph{JHEP} {\bf 10}
  (2008) 100}, [\href{http://arxiv.org/abs/0808.1094}{{\tt 0808.1094}}].

\bibitem{Lester:2007fq}
C.~Lester and A.~Barr, \emph{{MTGEN: Mass scale measurements in pair-production
  at colliders}},
  \href{http://dx.doi.org/10.1088/1126-6708/2007/12/102}{\emph{JHEP} {\bf 12}
  (2007) 102}, [\href{http://arxiv.org/abs/0708.1028}{{\tt 0708.1028}}].

\bibitem{Alwall:2009zu}
J.~Alwall, K.~Hiramatsu, M.~M. Nojiri and Y.~Shimizu, \emph{{Novel
  reconstruction technique for New Physics processes with initial state
  radiation}},
  \href{http://dx.doi.org/10.1103/PhysRevLett.103.151802}{\emph{Phys. Rev.
  Lett.} {\bf 103} (2009) 151802}, [\href{http://arxiv.org/abs/0905.1201}{{\tt
  0905.1201}}].

\bibitem{Baringer:2011nh}
P.~Baringer, K.~Kong, M.~McCaskey and D.~Noonan, \emph{{Revisiting
  Combinatorial Ambiguities at Hadron Colliders with $M_{T2}$}},
  \href{http://dx.doi.org/10.1007/JHEP10(2011)101}{\emph{JHEP} {\bf 10} (2011)
  101}, [\href{http://arxiv.org/abs/1109.1563}{{\tt 1109.1563}}].

\bibitem{Choi:2011ys}
K.~Choi, D.~Guadagnoli and C.~B. Park, \emph{{Reducing combinatorial
  uncertainties: A new technique based on MT2 variables}},
  \href{http://dx.doi.org/10.1007/JHEP11(2011)117}{\emph{JHEP} {\bf 11} (2011)
  117}, [\href{http://arxiv.org/abs/1109.2201}{{\tt 1109.2201}}].

\bibitem{Curtin:2011ng}
D.~Curtin, \emph{{Mixing It Up With MT2: Unbiased Mass Measurements at Hadron
  Colliders}}, \href{http://dx.doi.org/10.1103/PhysRevD.85.075004}{\emph{Phys.
  Rev.} {\bf D85} (2012) 075004}, [\href{http://arxiv.org/abs/1112.1095}{{\tt
  1112.1095}}].

\bibitem{Bai:2012gs}
Y.~Bai, H.-C. Cheng, J.~Gallicchio and J.~Gu, \emph{{Stop the Top Background of
  the Stop Search}},
  \href{http://dx.doi.org/10.1007/JHEP07(2012)110}{\emph{JHEP} {\bf 07} (2012)
  110}, [\href{http://arxiv.org/abs/1203.4813}{{\tt 1203.4813}}].

\bibitem{Giudice:2011ib}
G.~F. Giudice, B.~Gripaios and R.~Mahbubani, \emph{{Counting dark matter
  particles in LHC events}},
  \href{http://dx.doi.org/10.1103/PhysRevD.85.075019}{\emph{Phys. Rev.} {\bf
  D85} (2012) 075019}, [\href{http://arxiv.org/abs/1108.1800}{{\tt
  1108.1800}}].

\bibitem{Edelhauser:2012xb}
L.~Edelhauser, K.~T. Matchev and M.~Park, \emph{{Spin effects in the antler
  event topology at hadron colliders}},
  \href{http://dx.doi.org/10.1007/JHEP11(2012)006}{\emph{JHEP} {\bf 11} (2012)
  006}, [\href{http://arxiv.org/abs/1205.2054}{{\tt 1205.2054}}].

\bibitem{Lester:2014yga}
C.~G. Lester and B.~Nachman, \emph{{Bisection-based asymmetric M$_{T2}$
  computation: a higher precision calculator than existing symmetric methods}},
  \href{http://dx.doi.org/10.1007/JHEP03(2015)100}{\emph{JHEP} {\bf 03} (2015)
  100}, [\href{http://arxiv.org/abs/1411.4312}{{\tt 1411.4312}}].

\bibitem{James:1975dr}
F.~James and M.~Roos, \emph{{Minuit: A System for Function Minimization and
  Analysis of the Parameter Errors and Correlations}},
  \href{http://dx.doi.org/10.1016/0010-4655(75)90039-9}{\emph{Comput. Phys.
  Commun.} {\bf 10} (1975) 343--367}.

\bibitem{Cho:2008tj}
W.~S. Cho, K.~Choi, Y.~G. Kim and C.~B. Park, \emph{{M(T2)-assisted on-shell
  reconstruction of missing momenta and its application to spin measurement at
  the LHC}}, \href{http://dx.doi.org/10.1103/PhysRevD.79.031701}{\emph{Phys.
  Rev.} {\bf D79} (2009) 031701}, [\href{http://arxiv.org/abs/0810.4853}{{\tt
  0810.4853}}].

\bibitem{Choi:2009hn}
K.~Choi, S.~Choi, J.~S. Lee and C.~B. Park, \emph{{Reconstructing the Higgs
  boson in dileptonic W decays at hadron collider}},
  \href{http://dx.doi.org/10.1103/PhysRevD.80.073010}{\emph{Phys. Rev.} {\bf
  D80} (2009) 073010}, [\href{http://arxiv.org/abs/0908.0079}{{\tt
  0908.0079}}].

\bibitem{Park:2011uz}
C.~B. Park, \emph{{Reconstructing the heavy resonance at hadron colliders}},
  \href{http://dx.doi.org/10.1103/PhysRevD.84.096001}{\emph{Phys. Rev.} {\bf
  D84} (2011) 096001}, [\href{http://arxiv.org/abs/1106.6087}{{\tt
  1106.6087}}].

\bibitem{shlim:next}
S.~H. Lim, {work in progress}.

\end{thebibliography}\endgroup

\end{document}